\documentclass[prd,final,aps]{revtex4}

\usepackage{graphicx}
\usepackage{amssymb}

\begin{document}

%{\bf{\small version 4pm Feb 5, 2006}}

\bibliographystyle{prsty} % Choose Phys. Rev. style for bibliography

\title{The periodic standing-wave approximation: eigenspectral
computations for linear gravity and nonlinear toy models}

\author{Christopher Beetle}
\affiliation{Department of Physics,
Florida Atlantic University, Boca Raton, Florida 33431}  

\author{Benjamin Bromley}
\affiliation{Department of Physics,
University of Utah, Salt Lake City, Utah 84112}  
\author{Richard H.~Price} 
\affiliation{Department of Physics \& Astronomy and Center for 
Gravitational Wave Astronomy, University of Texas at Brownsville,
Brownsville, TX 78520}

\begin{abstract}
\begin{center}
{\bf Abstract}
\end{center}

The periodic standing wave approach to binary inspiral assumes rigid
rotation of gravitational fields and hence helically symmetric
solutions.  To exploit the symmetry, numerical computations must solve
for ``helical scalars,'' fields that are functions only of corotating
coordinates, the labels on the helical Killing trajectories.  Here we
present the formalism for describing linearized general relativity in
terms of helical scalars and we present solutions to the mixed partial
differential equations of the linearized gravity problem (and to a toy
nonlinear problem) using the adapted coordinates and numerical
techniques previously developed for scalar periodic standing wave
computations. We argue that the formalism developed may suffice for
periodic standing wave computations for  post-Minkowskian
computations and for full general relativity.
\end{abstract}

\maketitle

\section{Introduction}\label{sec:intro} 

The computational study of the inspiral of binary black holes is
important for the understanding of gravitational wave signals, and is
of inherent interest as a question in general relativity that can be
answered only with computation. It has therefore become the focus of
supercomputer codes that evolve Einstein's field equations forward in
time from initial conditions chosen to represent a starting
configuration of the inspiralling objects. The evolution codes,
however, typically become unstable on a timescale (set by the size of
the hole) short compared to a full orbit. Reliable calculations of the
final plunge are now feasible\cite{utbpuncture,goddardpuncture}, the
merger and ringdown of the final black hole fate of the system are
handled well with perturbation theory\cite{laz1}, and the early
inspiral is well approximated with post-Newtonian
computations\cite{PN}. What cannot be handled well is the intermediate
phase of the inspiral, the late epoch during which nonlinear effects
are too strong for a post-Newtonian approximation, but for which too
many orbits remain for stable numerical evolution.

It has long been recognized that the basis of an approximation scheme
should be the slow rate of inspiral, the small ratio of the orbital
time to the radiation damping time\cite{det,det2}.  Through an
adiabatic treatment of the slow inspiral, such an approximation could
give answers about the radiation and rate of inspiral in the
intermediate epoch.  In addition, when the rate of inspiral becomes
too rapid, the intermediate approximation could hand the problem off
to numerical evolution codes to do the final orbit or two and the
plunge. Along with the problem being handed off, would be the ideal
initial data for the subsequent evolution. The need for and the
concept of an intermediate approximation have been clear, but such an
approximation has not been easy to implement. Along with several
others\cite{WKP,WBLandP,rightapprox,paperI} we have based an approximation
for {\em slow} inspiral on  a numerical
computation of {\em no} inspiral. That is, we seek a numerical
solution of Einstein's equations for binary objects that are in
circular periodic motion, and whose ``helically symmetric'' fields
rotate rigidly with the source objects. (For a definition
of helical symmetry, see Sec.~II of Ref.~\cite{helicaldef}.)

The universality of gravitation suggests that the unchanging motion of
such a system is not compatible with outgoing radiation, and this
intuitive suggestion is confirmed by the mathematics of the theory.
We therefore seek a helically symmetric solution for the sources
coupled to standing waves, not to outgoing waves. In a linear theory,
standing waves, in the sense that we use the term, are a superposition
of half-ingoing and half-outgoing solutions.  From the fact that
the solution, in linear theory, is half the superposition of the ingoing
and outgoing solutions, one could extract the outgoing solution.  The crux
of our periodic standing wave (PSW) method is that even for highly
nonlinear binary inspiral fields there is an ``effective linearity.''
The standing wave solution, to good accuracy, is half the sum of the
outgoing plus ingoing solutions despite the nonlinearities. In general
relativity, therefore, our goal is to solve the standing wave
numerical problem and from that solution to extract an approximation
to the outgoing solution.

 It is important to understand why effective linearity can be correct
for inspiral. In the strong-field regions very close to the sources, 
the solution is  very insensitive to the choice of distant radiative
boundary conditions (ingoing, outgoing, standing wave). In this
near-source region a superposition of half the ingoing and half the
outgoing solution gives a good approximation solution, because 
it amounts to averaging two samples of the same thing. In the wave zone where
the outgoing and the ingoing solutions are very different, the fields
are weak enough that nonlinear effects are negligible, and once again
we can superpose. The  strong-field region and the 
boundary-influenced region should be widely separated unless the
sources are rotating very close to $c$, in which case the wave zone
will start just outside the sources. It is, however, not expected
that ultrarelativistic source motion can occur during the slow
inspiral epoch of motion.

We have recently\cite{eigenspec} been able to confirm effective
linearity.  This confirmation has been achieved with a model problem,
since the validity of effective linearity can only be carried out in a
model problem. In general relativity, there can be no ``true
outgoing'' solution available for confirmation until numerical
evolution codes are fully developed. In addition, the numerical
features of the helically symmetric standing wave calculation pose new
challenges very different from those of evolution codes, and are best
resolved in the simplest context possible.

The model problem in Ref.~\cite{eigenspec} used a nonlinear scalar
field theory with a pair of diametrically opposite point-like scalar
charges. The imposition of helical symmetry on the problem leads to a
boundary value problem for a system of mixed (hyperbolic and
elliptical) partial differential equations.  To solve that boundary
value problem efficiently, a set of numerical techniques was developed
that we called the ``eigenspectral'' method.  In the present paper we
report two important steps toward using the PSW method for full
general relativity: First, we develop the infrastructure for
describing gravitational fields.  In previous work with scalar toy
models helical symmetry was imposed by simply requiring that the
scalar field is a function only of three corotating coordinates
(labels on the Killing trajectories).  We find it useful to use the
expression ``helical scalars'' for such functions that depend only on
corotating coordinates. Our computation is done on a grid of
corotating coordinates, so straightforward computations can only be
carried out for helical scalars. But complications arise with tensor
fields. The components of a helically symmetric tensor field are
generally not helical scalars.  A resolution of this difficulty is to
compute only with projections of the tensor on a helically symmetric
basis, that is, on a basis that is Lie dragged by the helical
symmetry.  These projections would be helical scalars.  A helically
symmetric basis, however, has its own dynamics, and complicates the
time dependence of the projected fields.  In the infrastructure
developed here we show that the use of a corotating ``pure-spin''
basis leads to a remarkably simple description of helically symmetric
tensorial fields.  That infrastructure is presented in the explicit
context of linearized general relativity.

The second step taken in this paper is to present numerical results
showing that no new computational problems are encountered in dealing
with the helical scalars of the linearized gravity problem. Since no
new problems were anticipated this numerical work simply constitutes a
confirmation that the ``eigenspectral method,'' the set of techniques
developed for scalar models, appear to work equally well for
linearized gravity.  Those techniques include the use of (i)~``adapted
coordinates,'' a corotating coordinate system that conforms well to
the source regions and to the radiation field, (ii)~``multipole
filtering,'' the elimination of numerical noise associated with
angular differencing, by keeping only a few multipoles of the adapted
coordinates, and (iii)~the modification of the multipoles so that they
are computationally orthogonal at the level of machine precision.  In
this paper we provide numerical solutions using this set of
techniques. Since these solutions differ very little from the
numerical problems studied in detail in Ref.~\cite{eigenspec} the
presentation of linearized results is brief.  The solutions to
nonlinear problems are much more difficult than those for linearized
problems. Newton-Raphson iteration must be used, and convergence of
the iterative process has been the major challenge in numerical work.
Again, there is no apparent reason the problem should be more
difficult for the tensor-based helical scalars than for the nonlinear
scalar models of Ref.~\cite{eigenspec}. Nevertheless, it is an issue
worth checking, and initial results are briefly reported showing that
the scalar techniques work well for a model tensorial problem with a
simple toy nonlinearity.

The rest of this paper is organized as follows. Section
\ref{sec:scalar} gives a brief review of the scalar PSW problem in
order to introduce adapted coordinates and the application of
multipole filtering and the eigenspectral method in those
coordinates. Section \ref{sec:descrip} presents the description of the
helically symmetric fields of linearized gravity that is suitable for
computation. The field equations are given for general corotating
coordinates and series solutions are given, in corotating spherical
coordinates, for the problem of two equal masses in circular orbits
around each other. The field equations for linearized gravity using
adapted coordinates are given in Sec.~\ref{sec:adapcoord}. In that
section, also, are given the forms in adapted coordinates of other
elements of the computational problem, the inner boundary condtions
specifying the sources, and the outer boundary conditions specifying
the nature of radiation.  Section \ref{sec:numresults} presents
numerical results, comparing the series solutions of
Sec.~\ref{sec:descrip} with the solutions of the field equations using
the eigenspectral method.  In addition, in this section a description
is given of a toy nonlinear modification to linearized gravity, and
results are given demonstrating that the resulting ``theory'' can be
solved with the techniques that worked for the scalar case. The
implications for the next steps in the PSW program are discussed in
Sec.~\ref{sec:conc}. In particular, it is argued that almost all the
infrastructure for solving the PSW problems in the post-Minkowski
approximation and in full general relativity problem is probably
established in the work with the linearized problem. The Appendix
gives some detailed expressions needed for computations in adapted
coordinates.

Throughout the paper we use units in which $c=G=1
$ and we follow the conventions of Misner, Thorne and Wheeler\cite{MTW}.

\section{scalar models, coordinates, and numerics}\label{sec:scalar}
\subsection{Nonlinear scalar models}\label{subsec:nonlincal}

Here we review the model problem and numerical techniques of
Ref.~\cite{eigenspec}.
Our model problem was a nonlinear scalar field coupled to point-like
sources in Minkowski space, and satisfying the field equation
\begin{equation}\label{fieldtheory} 
\Psi_{;\alpha;\beta}g^{\alpha\beta}
+F=\nabla^2\Psi-\frac{1}{c^2
}\partial_t^2\Psi+ F={\rm Source}\,.
\end{equation}
In principle, the source was taken to be two points of unit scalar
charge in orbit around each other at angular frequency $\Omega$, and
at radius $a$. In practice, the computational problem used inner
boundary conditions on small, approximately spherical surfaces to
represent the effect of a point source; no source term was included in
the field equation that was computationally solved.
The velocity parameter for the system $v=a\Omega$ was taken
to be of order unity, representing the strong-field tight binary for
which post-Newtonian approximations are inadequate.  

The term $F$ contains the nonlinearity in our scalar model theory, and we
found the following form, with parameters $\lambda$ and $\Psi_0 $,
to be useful:
\begin{equation}\label{modelF} 
F=\frac{\lambda}{a^2}
\frac{\Psi^5}{\Psi_0^4+\Psi^4}\,.
\end{equation}
A crucial feature of $F $ is that, like the nonlinearities of general
relativity, it is very large near the sources, and becomes negligible
far from the sources.  The $\lambda$ multiplier allowed us to vary the
strength of the nonlinear term, and the $\Psi_0$ parameter allowed us
to vary the profile of the nonlinearity in the strong field region.  

Our scalar problem was defined by Eqs.~(\ref{fieldtheory}) and
(\ref{modelF}), and by the source motion at angular frequency $\Omega$
in the equatorial plane.  With standard spherical coordinates, helical
symmetry can be imposed on the solution $\Psi(t,r,\theta,\phi)$ by
restricting to solutions of the form $\Psi(r,\theta,\varphi) $, where
$\varphi $ is the comoving azimuthal coordinate $\phi-\Omega t $.  By
restricting the solution in this way, we have eliminated the
possibility of ``evolution.''  For such helically symmetric solutions
a change in time by $\Delta t$ is the same as a change the azimuthal
angle $\Delta\phi=-\Omega\Delta t$.

When the restriction to helical symmetry is made, the 
field equation becomes
\begin{displaymath}
{\cal L}\Psi\equiv 
\frac{1}{r^2}\frac{\partial}{\partial r}
\left(r^2\frac{\partial\Psi}{\partial r}\right)
+\frac{1}{r^2\sin\theta}\frac{\partial}{\partial\theta}
\left(\sin\theta\frac{\partial\Psi}{\partial\theta}\right)\hspace{1.5in}
\end{displaymath}
\begin{equation}\label{ourL} 
\hspace{.5in}+\left[\frac{1}{r^2\sin^2\theta}
-{\Omega^2}
\right]\,\frac{\partial^2\Psi}{\partial\varphi^2}
={\rm Source}-F(\Psi)\equiv\sigma(\Psi)\,,
\end{equation}
and the mixed nature of the partial differential equation becomes
obvious.  The principal part of this quasilinear equation is elliptic
inside a cylinder at $r\sin\theta=1/\Omega $, and hyperbolic outside
that cylinder. The problem is to be solved with radiative conditions
(ingoing, outgoing, or standing wave as described below) on a
spherical surface at large distances from the sources.  Well posed
problems in physics typically supply cauchy data on open surfaces to
hyperbolic equations, and Dirichlet or Neumann data on closed surfaces
to elliptic equations.  Our model is unusual in that it leads to a
boundary value problem with ``radiation'' conditions on a closed
surface surrounding a mixed problem. Though unusual, our problem is
intuitively well posed, and passes a computational test: we have found
no fundamental difficulty in solving models of this type
numerically. Furthermore, a careful analysis\cite{torre} proves that
solutions exist and are stable for a closely related problem.

``Standing wave'' solutions -- half ingoing and half outgoing -- are at
the heart of our method, but there is not an unambiguous definition of
standing wave solutions in a nonlinear theory. Our procedure is to
find the outgoing ${\cal L}^{-1}_{\rm out}$ and ingoing ${\cal
L}^{-1}_{\rm in}$ Green functions for Eq.~(\ref{ourL}).  In principle, we
can then iterate to find a solution of Eq.~(\ref{ourL}). The iteration
\begin{equation}\label{outit} 
\Psi^{(n+1)}_{\rm out}={\cal L}^{-1}_{\rm out}\big[\sigma(\Psi^{(n)}_{\rm out}
)\big]\,,
\end{equation}
if it converges, gives $\Psi_{\rm out}$,
our nonlinear outgoing solution (and similarly for $\Psi_{\rm in}
$), while  
the convergent result of 
\begin{equation}\label{stdit} 
\Psi^{(n+1)}_{\rm std}
={\textstyle\frac{1}{2}} \left({\cal L}^{-1}_{\rm out}+{\cal
L}^{-1}_{\rm in} \right) \big[\sigma(\Psi^{(n)}_{\rm std} )\big]\,
\end{equation}
is what we mean by our nonlinear standing wave solution
$\Psi_{\rm std}$. The standing wave solution
$\Psi_{\rm std}$
is fundamentally different from 
$(\Psi_{\rm out}+\Psi_{\rm in})/2$, but if effective linearity
is correct, the two  are very nearly equal.
(Note: In practice, for strong nonlinearities, the direct
iteration described above must be replaced by Newton-Raphson
iteration.)

A central idea of the PSW approximation is that the ``exact'' (i.e.\,,
numerical) solution to the standing wave problem is an excellent
approximation to half the sum of the outgoing and ingoing
solutions. If this is so, it means that from the standing wave
solution we should be able to extract an excellent approximation to
the outgoing and the ingoing solutions. (It should be noted that this
statement is meaningful for nonlinear model field theories in which
there is meaning to an outgoing helically symmetric or ingoing
helically symmetric solution to the field theory. For general
relativity, a helically symmetric spacetime with outgoing radiation is
impossible.  As explained in Ref.~\cite{eigenspec}, the relevance and
justification of the PSW approximation for full general relativity lie
in the fact that the method gives an approximation only for a limited
region of spacetime.)

The extraction of the approximate nonlinear outgoing solution from the
computed standing wave solution is a direct application of the
concepts underlying the argument for effective linearity. In the weak
wave zone, far from the sources, the field theory is very nearly
linear and the solution must be very nearly a standing wave solution
to that linear theory. It is, therefore, straightforward to
deconstruct it into outgoing and ingoing solutions. The extracted
outgoing solution can be continued inward through the induction zone
into the near-field zone.  If the theory were completely linear, this
continuation would be valid up to the source points. Due to the
nonlinearity, however, this procedure is no longer valid at distances
so close to the source points that the nonlinearity of the theory is
important.

In the near field region close to the source points, at a distance
small compared to a wavelength, the solution is highly insensitive to
the nature of the boundary conditions (i.e.\,, whether they are
outgoing or ingoing). Here, we can use the standing wave solution
itself as the outgoing solution.  This near field solution should
extend out to the weak field region and overlap with the extracted
outgoing solution described above.  These two solutions, the weak
field outgoing solution outside the strong field region, and the
standing wave solution in the strong field region, are then patched
together, with some blending in the region of overlap, and the result
is the PSW extracted approximation to the nonlinear outgoing solution.

\subsection{Adapted coordinates}
%

%%%%%%%%%%%%%%%%%%%%%%%%%%%%%%%%%%%%%%%%%%%%%%
\begin{figure}[ht]   %FIG 1 
\includegraphics[width=.26\textwidth]{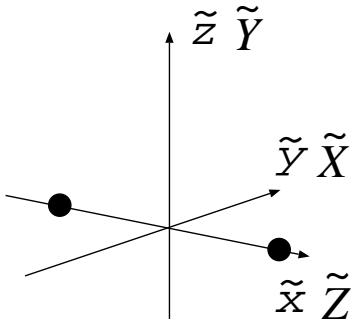}

\caption{Two systems of corotating Cartesian 
coordinates.
\label{fig:XYZ}}
\end{figure}
%%%%%%%%%%%%%%%%%%%%%%%%%%%%%%%%%%%%%%%%%%%%%%

It is useful to identify a number of coordinate systems to describe
the physical problem, including three Cartesian systems. The system
$\{x,y,z\}$ is that of inertial Cartesian system in which the $z$ axis is
the rotation axis for the source objects, with rotation in the
positive sense about the $z$ axis. In general we use tildes to
distinguish the corotating version of a quantity 
when that quantity occurs in forms both related to inertial and to
corotating systems.  The set
$\{\widetilde{x},\widetilde{y},\widetilde{z}\}$, in this sense, is the
corotating set of Cartesian coordinates for which $\widetilde{z}=z $
and for which the source points remain fixed on the $\widetilde{x}$ axis. The
system $\{r,\theta,\phi\}$ is that of inertial spherical coordinates defined
in the usual way relative to $\{x,y,z\}$. The system
$\{r,\theta,\varphi\}$ is a set of coroting spherical polar
coordinates, defined by the usual transformation relative
to $\{\widetilde{x},\widetilde{y},\widetilde{z}\}$ . The two systems
of spherical coordinates are related by $\varphi=\phi-\Omega t$.  The
Cartesian system $\{\widetilde{X},\widetilde{Y},\widetilde{Z}\}$ is a
convenient renaming of
$\{\widetilde{x},\widetilde{y},\widetilde{z}\}$, with $\widetilde{Y}$
the rotational axis, and $\widetilde{Z}$, the axis through the source
points.

Our adapted coordinate system $\{\chi,\Theta,\Phi\}$ is a corotating
two-center bipolar coordinate system defined, relative to the
$\{\widetilde{X},\widetilde{Y},\widetilde{Z}\}$ system, by
\begin{eqnarray}
\chi&\equiv&
\left\{\left[\left(\widetilde{Z}-a\right)^2+\widetilde{X}^2
+\widetilde{Y}^2\right]
\left[\left(\widetilde{Z}+a\right)^2+\widetilde{X}^2
+\widetilde{Y}^2\right]\right\}^{1/4}\label{chiofXYZ}\\
\Theta&\equiv&
\frac{1}{2}\tan^{-1}\left(\frac{2\widetilde{Z}
\sqrt{\widetilde{X}^2+
\widetilde{Y}^2
\;}}{
\widetilde{Z}^2-a^2-\widetilde{X}^2-\widetilde{Y}^2
}\right)\label{ThetofXYZ}\\
\Phi&\equiv&\tan^{-1}{\left(\widetilde{Y}/\widetilde{X}\right)}
\label{PhiofXYZ}\,,
\end{eqnarray}
where $a $ is the distance from the center to each of the source
points.

\begin{figure}[ht] %%%%%%FIG 2
\includegraphics[width=.26\textwidth]{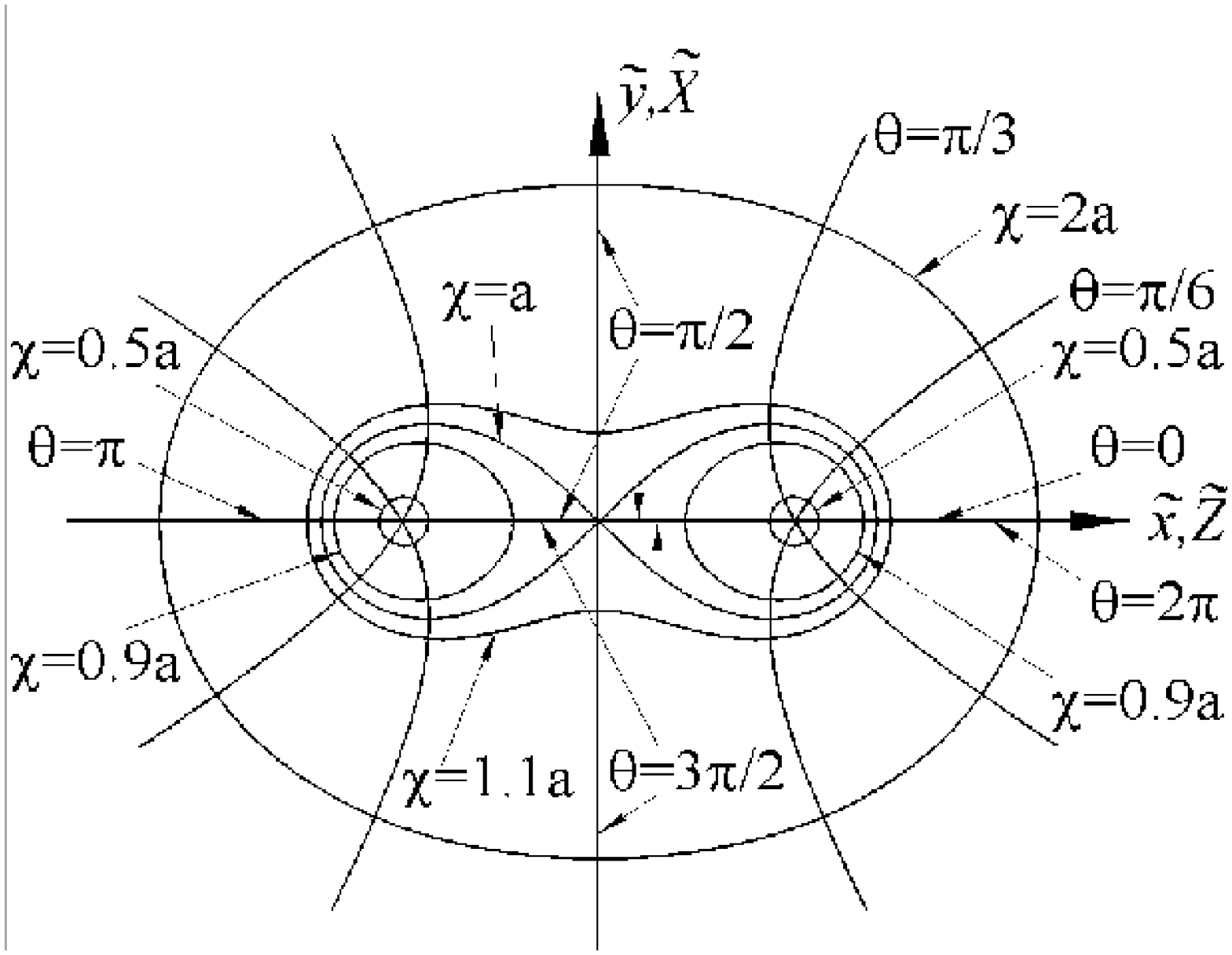}\quad\quad
\includegraphics[width=.26\textwidth]{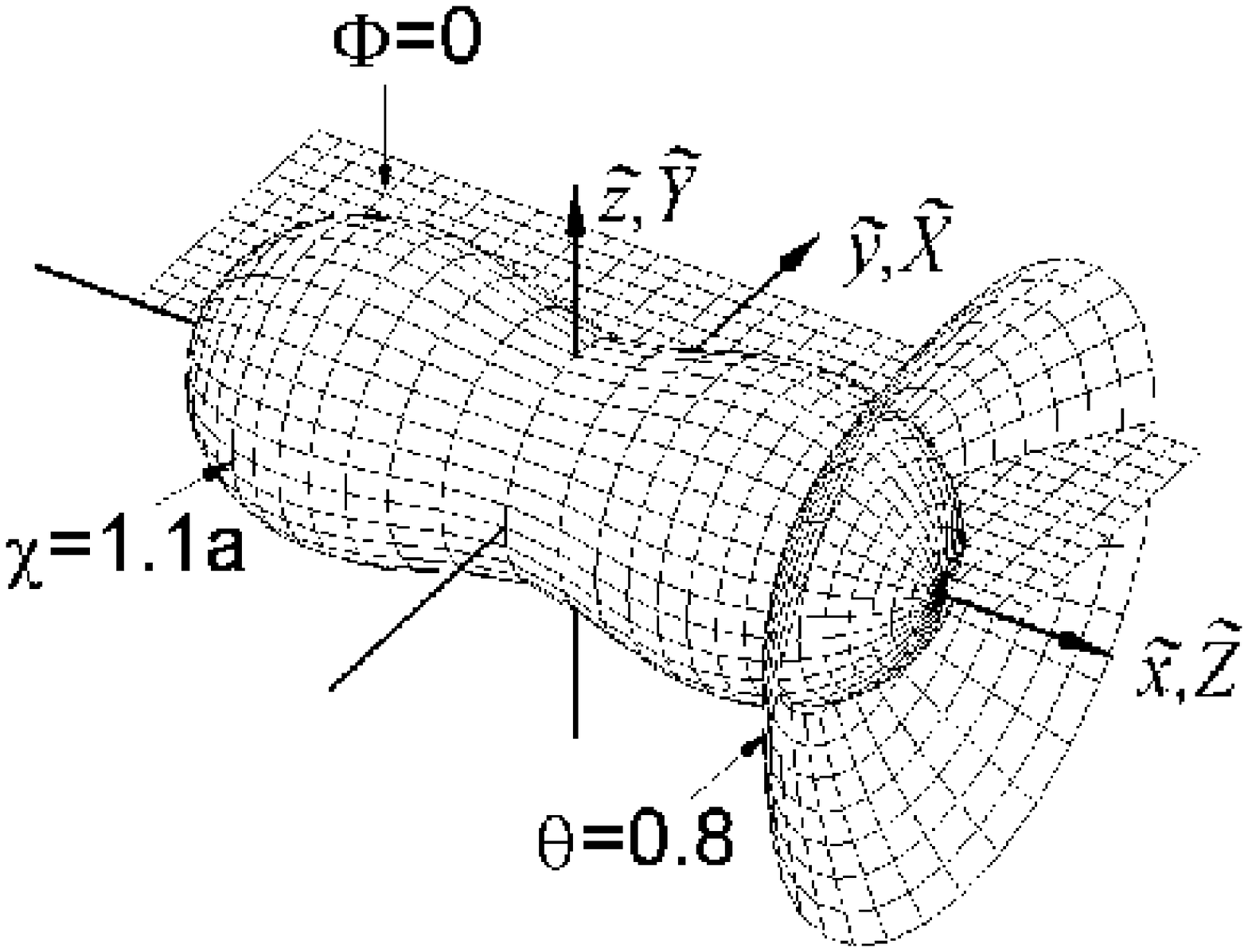} \caption{Two-center
bipolar adapted coordinates. On the left is shown curves of
coordinates $\chi$ and $\Theta $ in the $\Phi=0$ orbital plane. The
surfaces on the right show surfaces of constant $\chi $, $\Theta$, and
$\Phi$. Here $a$ is the distance from either of the two centers
(locations of the point sources) to the midpoint between the centers.
The $\widetilde{x},\widetilde{y},\widetilde{z}$ coordinate system is a
corotating Cartesian system for which the $\widetilde{z} $ axis is the
rotation axis; for the $\widetilde{X},\widetilde{Y},\widetilde{Z}$ a
corotating Cartesian system the $\widetilde{Z} $ axis is the line
through the centers.  \label{fig:2dros}}
\end{figure}

For $\chi\gg a$, the adapted coordinates become a corotating system of
spherical coordinates defined relative to
$\{\widetilde{X},\widetilde{Y},\widetilde{Z}\}$.  (That is, the
$\widetilde{Z}$ axis is the azimuthal axis for the $\Phi $
coordinate.) The adapted coordinates, pictured in
Fig.~\ref{fig:2dros}, are discussed in greater detail in Ref.~\cite{eigenspec}.
%%%%%%%%%%%%%%%%%%%%%%%%%%%%%%%%%%%%%%%%%%%%%%
%%%%%%%%%%%%%%%%%%%%%%%%%%%%%%%%%%%%%%%%%%%%%%

We originally\cite{paperI} solved the model problem of
Eq.~(\ref{ourL}) in the $\{r,\theta,\varphi\}$ system with
more-or-less straightforward finite differencing and direct matrix
inversion. (The mixed nature of the partial differential equations
prevents the use of such efficient techniques as overrelaxaton.)
This approach was successful (iterations converged) for models with a
limited range of source velocities and nonlinearities. Subsequently we
developed an innovative numerical method that gave remarkably good
results for the scalar problem, with very little computational
cost. The new method is based on three elements.

First, we used the adapted coordinates discussed above.  The solutions
to the finite difference form of Eq.~(\ref{fieldtheory}) in these
coordinates turned out to be plagued by what appeared to be angular
noise. This noise was eliminated with multipole filering, a form of
smoothing of the angular variation. In this method the scalar field
was expanded in spherical harmonics of the angular functions $\Theta$
and $\Phi$.  The fact that the adapted coordinates are well suited
both to the source structure and to the radiation field suggests that
good accuracy can be achieved when only a few multipoles are kept.  In
fact, good accuracy was found when only the monopole and quadrupole
were kept in the case of sources speeds around 0.3$c$ -0.4$c$ or less. At
higher speeds the radiation field develops sharper gradients and more
multipoles must be kept to achieve reasonable accuracy.

\subsection{inner and outer boundary conditions}
For identical point sources of unit scalar charge moving in the
equatorial plane at angular velocity $\Omega$ in circular orbits of
radius $a$, the source used in Eq.~(\ref{fieldtheory}) was
\begin{equation}\label{ptsource} 
S=-\,\gamma^{-1}
\frac{\delta(r-a)}{a^2}
\;\delta(\theta-\pi/2)\left[\delta(\varphi)+\delta(\varphi-\pi)
\right]\,,
\end{equation}
with $\gamma\equiv1/\sqrt{1-v^2\;}=1/\sqrt{1-a^2\Omega^2 \;}$.  For
this source, it was shown in Ref.~\cite{eigenspec} that the small-$\chi$
limit of $\Phi $ is
\begin{equation}\label{innerscalbc} 
\Psi=\frac{1}{4\pi}\frac{2a}{\chi^2
}\,\frac{1}{\sqrt{1+\gamma^2v^2
\sin^22\Theta\cos^2\Phi
\;}}\,.
\end{equation}
We take this as an inner boundary condition at a small value of $\chi
$ that determines one limit of our computational grid.

The outer boundary condition used is simply 
the ingoing or outgoing  Sommerfeld condition expressed 
in terms of the adapted coordinates. In Ref.~\cite{eigenspec}
it was shown that in adapted coordinates this condition becomes
\begin{equation}\label{FDMoutbcapprox} 
\frac{1}{\chi}\frac{\partial}{\partial\chi}\left(\chi\Psi\right)
=\pm\Omega\,\left(\cos\Phi\frac{\partial\Psi}{\partial\Theta}
-\frac{\cos\Theta}{\sin\Theta}\,\sin\Phi
\frac{\partial\Psi}{\partial\Phi}
\right) \,,
\end{equation}
in which the upper and lower signs correspond, respectively, to the
outgoing and ingoing conditions.

\subsection{multipole filtering  and the eigenspectral technique}

In practice, multipole filtering was carried out as follows.  If there
are $N $ grid locations $\Theta_i$, $\Phi_j$, of the angular
coordinates $\Theta $, $\Phi$, then at each value of the radial
coordinate $\chi$ there are $N$ values of the scalar field
$\Psi(\chi,\Theta_i,\Phi_j)$. In multipole filtering, a set of $M\leq
N $ spherical harmonics $Y_{ij}$ is used, at each value of $\chi$, as
weighting factors for these scalar field values, and $M$ weighted sums
(i.e.\,, multipole projections) are taken of the scalar field values. In the
same manner, the $N$ field equations at $\chi$ are projected into $M$
sums.  Those $M$ projected equations are then solved for the $M$
projections of the field.

It was found that this procedure did not work if the multipole weights
$Y_{ij}$ were simply taken to be $Y_{Lm}(\Theta_i ,\Phi_j )$, the
continuum spherical harmonics evaluated at the discrete angular grid
locations. These projection weights are only orthogonal in the
continuum limit, and their failure to be numerically orthogonal to
high precision was the probable source of angular noise that plagued
computations. Slightly modified weights, in place of the grid-evaluated
spherical harmonics, gave us weights that was orthogonal to
the level of machine precision. The use of these modified weights in
multipole filtering eliminated the problem of angular noise, while at
the same time significantly reducing the number of equations to be
solved, and hence reducing the computational burden. See
Ref.~\cite{eigenspec} for details.

\begin{figure}[ht] %%%%FIG 3
\includegraphics[width=.23\textwidth]{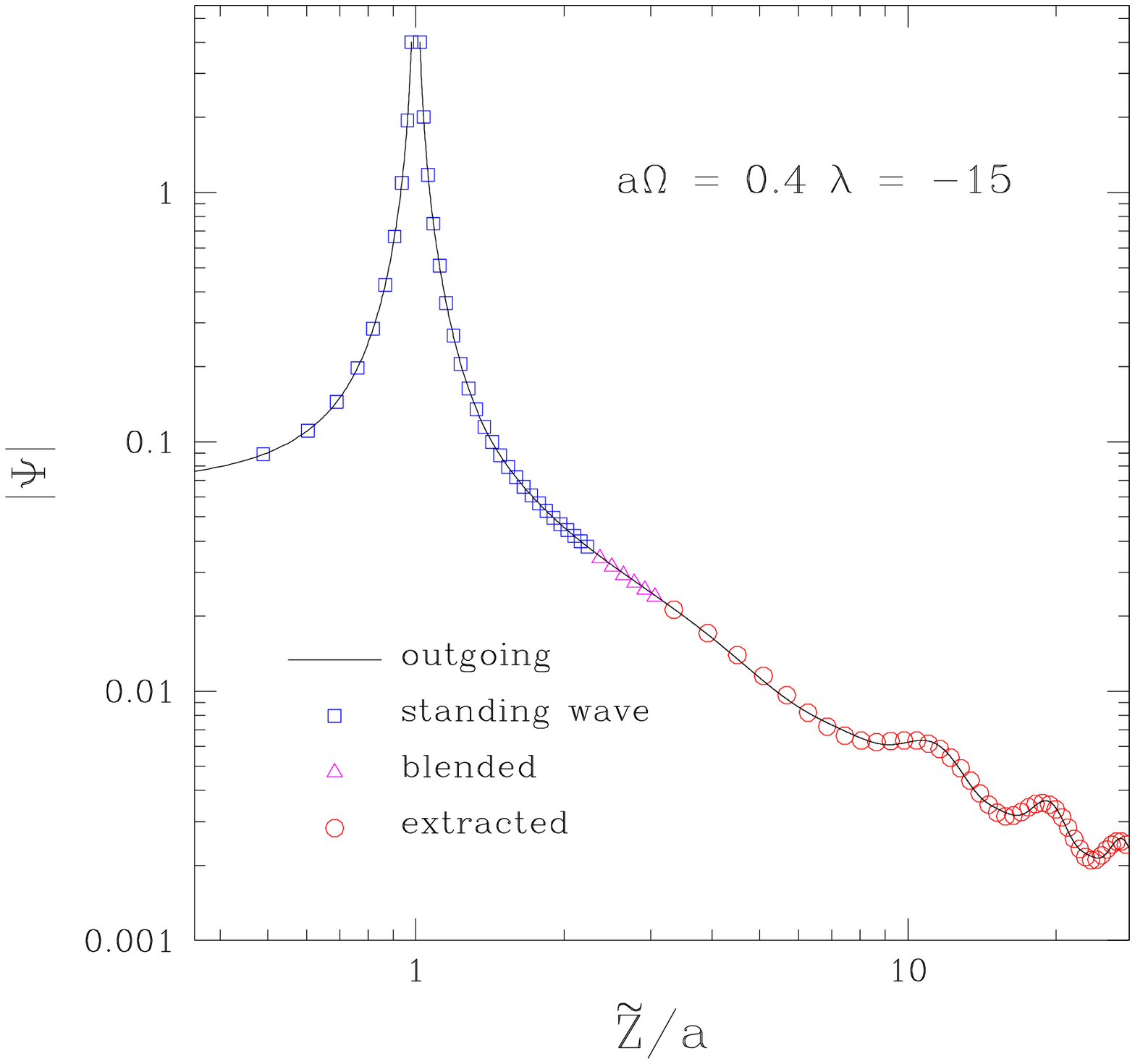}
\includegraphics[width=.23\textwidth]{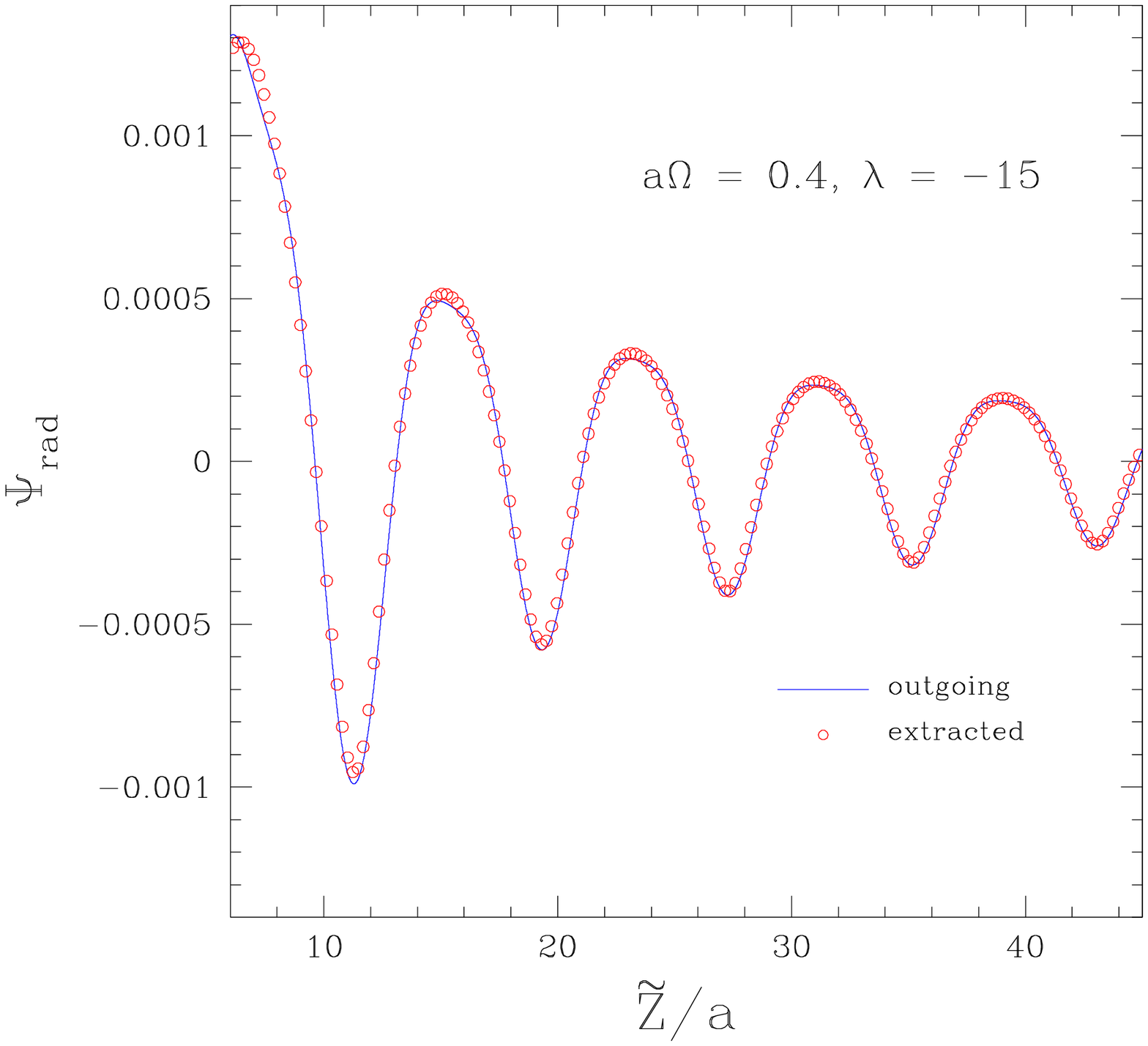} \caption{ The computed
nonlinear outgoing solution compared with an approximate outgoing
solution extracted from the computed nonlinear standing wave
solutions. The solid curves show the computed outgoing solution for a
scalar model with nonlinearity parameters $\lambda=-15$ and
$\Psi_0=-0.15$ and with source velocity $a\Omega=0.4c $.  The
data-like markers show the approximate outgoing solution extracted
from the standing wave solution; the shape of the markers indicate
whether the extracted solution corresponds to the deconstruction of
the linearized solution, the standing wave solution, or a region of
blending of the linearized outgoing and standing wave solutions.  The
computations of both the standing wave solution and the outgoing
solution were carried out using quadrant symmetry with a $40\times80$
angular grid, and with 16001 divisions in $\chi$ ranging from
$\chi=0.02a $ to $80a $.  The multipole filtering kept only the
multipoles corresponding to $\ell=0,2$ and 4.  \label{fig:extract}}
\end{figure}

We focused, in Ref.~\cite{eigenspec}, on the most important question
that can be answered with these models and numerical methods: Does
effective linearity work? Can we extract a good approximation to the
outgoing nonlinear problem from the sort of standing wave computation
we will be limited to when dealing with Einstein's theory?
Figure~\ref{fig:extract} gives strong evidence that we can.  In that
figure, the computed outgoing nonlinear solution is shown as a solid
curve. The data-type points represent the outgoing solution extracted
in the manner described above in Sec.~\ref{subsec:nonlincal}.  For the
parameters $\lambda=-15$ and $\Psi_0=0.15$ in Fig.~\ref{fig:extract},
nonlinearities are significant, strong enough to reduce field strength
by around two-thirds. The outgoing and standing wave solutions were
each computed by the Newton-Raphson version of the iteration in
Eqs.~(\ref{outit}),(\ref{stdit}).  We have run models with much
stronger nonlinearity and have found equally good agreement of the
true outgoing solution and the extracted approximation. The validity
of effective linearity should, in fact, become questionable not for
stronger nonlinearity, but only for physically implausible high source
velocity.

In addition to confirming effective linearity, computation with the
 model has also allowed some early insights about sensitivity to
 source details. By varying the multipole content of the inner boundary data
 we  explored the impact of source structure on the radiation
 field. The result (detailed in Ref.~\cite{eigenspec}) is in perfect
 accord with physical intuition; the radiation is insensitive to
 source structure unless the source size becomes comparable to the
 separation of the sources (i.e.\,, unless the moments ascribable to
 the structure of the individual sources are comparable to the
 quadrupole moment due to the separation of the mass points). The
 equivalent question for Einstein's theory is more difficult, but we
 should eventually be able to give clear quantitative answers.

%%%%%%%%%%%%%%%%%%%%%%%%%%%%%%%%%%%%%%%%%%%%%%
%%%%%%%%%%%%%%%%%%%%%%%%%%%%%%%%%%%%%%%%%%%%%%

%
\section{the description of helically symmetric linearized
gravitational fields}\label{sec:descrip} 

\subsection{The physical problem}
We use the standard
description and notation for the linearization of Einstein's field
equations (see Chap.~8 of Ref.~\cite{MTW}). The perturbations from flat spacetime
$h_{\mu\nu}$ are defined by 
\begin{equation}
g_{\mu\nu}=\eta_{\mu\nu}+h_{\mu\nu}\,,
\end{equation}
and the trace-reversed perturbations 
$\bar{h}_{\mu\nu}$ are defined by 
\begin{equation}
\bar{h}_{\mu\nu}\equiv {h}_{\mu\nu}-\frac{1}{2}\eta_{\mu\nu}h\,,
\end{equation}
where $h\equiv\eta^{\mu\nu}{h}_{\mu\nu}$. For linearized computations, 
$\bar{h}_{\mu\nu}$ is treated as a tensorial field in Minkowski spacetime,
and indices are raised and lowered with the Minkowskian metric $\eta_{\mu\nu}$.
In the gravitational Lorentz gauge
\begin{equation}
\bar{h}^{\mu\nu}_{,\mu}
=0\,,
\end{equation}
the linearized  field equations of general relativity are
\begin{equation}\label{boxhbareqT} 
\bar{h}_{\mu\nu,\alpha}^{\ \ \ \ ,\alpha}=-16\pi T_{\mu\nu}\,.
\end{equation}

Our physical problem is that of two points, each of mass $m_0$, in
circular orbits , with radius $a$, angular velocity $\Omega$, and
hence speed $v=a\Omega $ through the background Minkowski
spacetime.
The general form of the stress-energy source for linearized
theory\cite{zerilli70} is
\begin{equation}\label{genformTmunu} 
T^{\mu\nu}=m_0\int_{-\infty}^{\infty}
\delta^{(4)}(x^\alpha-z^\alpha(\tau))\,\frac{dz^\mu}{d\tau}
\,\frac{dz^\nu}{d\tau}\;d\tau=m_0\frac{U^\mu U^\nu}{U^0}
\frac{\delta(r-R(t))}{r^2}\,\delta^{(2)}(\Omega-\Omega(t))\,,
\end{equation}
in which 
\begin{equation}\label{U0Uphi}
U^0\equiv\gamma=1/\sqrt{1-v^2\;}
\quad\quad  
U^x=\mp v\gamma\sin\Omega t\quad\quad
U^y=\pm v\gamma\cos\Omega t\,.
\end{equation}
The signs of $U^x$ and $U^y$ are different for the two source
particles. One, call it particle 1, is at $\phi=\Omega t$; the other,
particle 2, is at $\phi=\Omega t+\pi$. In Eq.~(\ref{U0Uphi}) the upper
sign corresponds to particle 1, the lower to particle 2.  The explicit
nonvanishing components, in the inertial $t,r,\theta,\phi $ system,
are then,
\begin{equation}\label{T00value} 
T^{tt}=m_0\,
\gamma
\;\frac{\delta\left(r-a\right)}{a^2}
\delta\left(\theta-\pi/2 \right)
\left[
\delta\left(\varphi\right)+\delta\left(\varphi-\pi\right)
\right]
\end{equation}
\begin{equation}\label{T0jval}
T^{tx}=\mp v\sin{\Omega t}\;T^{tt}
\quad\quad
T^{ty}=\pm v\cos{\Omega t}\;T^{tt}
\end{equation}
\begin{equation}\label{Tijvalue} 
T^{xx}=v^2\sin^2{\Omega t}\;T^{tt}
\quad\quad
T^{yy}=v^2\cos^2{\Omega t}\;T^{tt}
\quad\quad
T^{xy}=-v^2\sin{\Omega t} \cos{\Omega t}\;T^{tt}\,.
\end{equation}

We use this helically symmetric stress energy as an explicit source
when we derive the series solutions to Eq.~(\ref{boxhbareqT}). 
For the solution of Eq.~(\ref{boxhbareqT}) via the eigenpectral 
method of Ref.~\cite{eigenspec}, however, 
we find $\bar{h}_{\alpha\beta} $ in the limit of small distance from
the particles, and impose this solution as inner boundary conditions
on the homogeneous form of Eq.~(\ref{boxhbareqT}).  To get the
near-particle solution we use the Li\'enard-Wiechert solution of
Eq.~(\ref{boxhbareqT})
\begin{equation}\label{LWsoln}
\left.\bar{h}^{\alpha\beta}(\vec{x})=-\,4m_0
\frac{U^{\alpha}(\tau)U^{\beta}(\tau)
}{\vec{U}\cdot\left(\vec{x}
-\vec{r}(\tau)\right)}\right|_{\rm ret}\ .
\end{equation}
The retardation condition ``ret,'' means that $\tau$ is to be evaluated at 
time such that 
\begin{equation}
|\vec{x}-\vec{r}(\tau)|=0.
\end{equation}

For the particle at $\varphi=0$, we now evaluate this approximately by
assuming that the particle moves with constant velocity $x=a $,
$y=a\Omega t_{\rm part}=vt_{\rm part}$, $z=0 $, through the inertial coordinate
frame.
At field point $t,x,y,z$ we have the following retardation condition 
for the particle at $x=a
$:
\begin{equation}
0=-(t-t_{\rm part})^2
+(x-a)^2+(y-vt_{\rm part}
)^2+z^2\,,
\end{equation}
with the solution 
\begin{equation}\label{tpart1} 
t_{\rm part}=\gamma^2\left[
t-yv-\sqrt{(t-vy)^2+(r^2-t^2)/\gamma^2
\;}\;
\right]
\end{equation}
in which
\begin{equation}
r^2\equiv
(x-a)^2+z^2+y^2\ .
\end{equation}
To express this in corotating coordinates, we next introduce the
 notations and approximations
\begin{equation}
y=\widetilde{y}+vt\quad\quad
\widetilde{x}=x\quad\quad\widetilde{z}=z\quad\quad\
\widetilde{r}^2\equiv(\widetilde{x}-a)^2+\widetilde{y}^2+\widetilde{z}^2\,.
\end{equation}
With these, Eq.~(\ref{tpart1}) simplifies to
\begin{equation}
t_{\rm part}=t-v\gamma^2\widetilde{y}-\gamma
\sqrt{\widetilde{r}^2+\gamma^2v^2\widetilde{y}^2\;}\,,
\end{equation}
and, finally,  we can evaluate 
\begin{displaymath}
\left.{\vec{U}\cdot\left(\vec{x}
-\vec{r}(\tau)\right)}\right|_{\rm ret}
=-U^0(t-t_{\rm part})+U^y(y-v t_{\rm part})
\end{displaymath}
\begin{displaymath}
=-\gamma\left[t-t_{\rm part}-v(y-vt_{\rm part})\right]
\end{displaymath}
\begin{equation}\label{finretard}
=-\,\gamma^{-1}
\left(t-t_{\rm part}\right)
+v\gamma\widetilde{y}
=-\,\sqrt{\widetilde{r}^2+\gamma^2v^2\widetilde{y}^2
\;}\ .
\end{equation}

With Eqs.~(\ref{U0Uphi}) and (\ref{finretard}), the expression 
in Eq.~(\ref{LWsoln}) gives us the inner boundary conditions to 
be used for solving 
Eq.~(\ref{boxhbareqT}):
\begin{eqnarray}
\bar{h}^{nn}&=&4m_0\frac{\gamma^2}
{\sqrt{\widetilde{r}^2+\gamma^2v^2\widetilde{y}^2\;}}
\label{barhinnerfirst}  \\
\bar{h}^{nx}=\bar{h}^{xn}&=&4m_0\frac{\gamma^2}
{\sqrt{\widetilde{r}^2+\gamma^2v^2\widetilde{y}^2\;}}
\;\left[\mp v\sin{\Omega t}
\right]\\
\bar{h}^{ny}=\bar{h}^{yn}&=&4m_0\frac{\gamma^2}
{\sqrt{\widetilde{r}^2+\gamma^2v^2\widetilde{y}^2\;}}
\;\left[\pm v\cos{\Omega t}\right]\\
\bar{h}^{xx}&=&4m_0\frac{\gamma^2}
{\sqrt{\widetilde{r}^2+\gamma^2v^2\widetilde{y}^2\;}}
\;\left[v^2
\sin^2{\Omega t}\right]\\
\bar{h}^{yy}&=&4m_0\frac{\gamma^2}
{\sqrt{\widetilde{r}^2+\gamma^2v^2\widetilde{y}^2\;}}
\;\left[v^2
\cos^2{\Omega t}\right]\\
\bar{h}^{xy}=\bar{h}^{yx}&=&4m_0\frac{\gamma^2}
{\sqrt{\widetilde{r}^2+\gamma^2v^2\widetilde{y}^2\;}}
\;\left[-v^2\cos{\Omega t}\sin{\Omega t}\right]\,.\label{barhinnerlast}
\end{eqnarray}

The outer boundary conditions, roughly speaking, are the conditions
that nonradiative parts of the field fall off as $1/r^n$, and the
radiative parts of the field satisfy simple ingoing or outgoing
Sommerfeld conditions.  The details are given after the presentation
of the formalism for describing helically symmetric tensorial fields.

\subsection{Description of helically symmetric tensorial fields}

Imposing helical symmetry on the tensorial field 
means that the Lie derivative of  $\bar{h}_{\mu\nu}$
vanishes along the helical Killing field, or
\begin{equation}
{\cal L}_{\xi}\left(\bar{h}_{\mu\nu}\right)=0\,.
\end{equation}
Here the helical Killing vector is
\begin{equation}
\xi=\partial_t+\Omega\partial_\phi=\partial_t.
\end{equation}
The first expression gives $\xi$ in terms of the inertial spherical
background coordinates $t,r,\theta,\phi$, and the second in terms of
the corotating spherical background coordinates
$t,r,\theta,\varphi=\phi-\Omega t$.  As explained in
Sec.~\ref{sec:intro}, our computational unknowns are fields on a grid
of corotating coordinates. We must therefore cast the field equations
in terms of a set of functions that are ``helical scalars,'' functions
only of the spatial corotating Minkowski coordinates such as
$\{r,\theta,\varphi\}$
$\{\widetilde{x},\widetilde{y},\widetilde{x}\}$,
$\{\widetilde{X},\widetilde{Y},\widetilde{Z}\}$, or the adapted
coordinates $\{\chi,\Theta,\Phi\}$.

The key to this is to project on a corotating basis. A corotating
vector $\widetilde{\bf V}$ is one for which ${\cal
L}_\xi(\widetilde{\bf V})=0 $.  If $\widetilde{\bf V}$ and
$\widetilde{\bf W}$ are both corotating vectors, and
$\bar{h}_{\mu\nu}$ is helically symmetric, then $h_{\mu\nu}V^\mu
W^\nu$ is a helically symmetric scalar, a function only of three
corotating coordinates.  Our approach, then is to use a coroating
basis and to project all components of $h_{\mu\nu}$ on this basis.

If not done with some care, projection on a corotating basis, can
destroy the simplicity of the linearized field equations
(\ref{boxhbareqT}). 
Since the inertial basis is
covariantly constant, the equations separate for 
the components of $\bar{h}_{\mu\nu}$; each component satisfies 
its own equation, and the system of equations separates into a set of
scalar-like equations. This is not true in general for the scalars
formed by projections with the corotating bases. Choices can be made,
however, that result in a high degree of separation, and a very simple
set of equations.

To achieve this simplicity we start by defining the covariantly
constant orthonormal inertial basis system for the Minkowski
background:
\begin{equation}
{\bf n}=\partial_t\quad
{\bf e}_x=\partial_x\quad
{\bf e}_y=\partial_y\quad
{\bf e}_z=\partial_z\quad\,.
\end{equation}
Here we closely follow the analysis given by
Thorne\cite{kiprevmodphys}. That analysis uses the method of Mathews
\cite{mathews}, and introduces a set of second rank symmetric spatial
basis tensors ${\bf t}_{2,j}$ of spin 2, i.e.\,, tensors that
transform among themselves as an irreducible represenation of the
rotation group of order 2. To these we add an additional spin 0
spatial tensor ${\bf t}_{0,0}$ and the spin 0 and spin 1 second-rank
symmetric tensors needed to include the timelike direction
\begin{eqnarray}
{\bf t}_{nn}&\equiv&{\bf n}{\bf n}\label{tnndef}  \\
{\bf t}_{n0}&\equiv&{\bf n}{\bf e}_z\\
{\bf t}_{n,\pm1}&\equiv&
%\textstyle{\frac{1}{\sqrt{2\;}}}
%\left[{\bf n}{\bf s}_{\pm1}+{\bf s}_{\pm1}{\bf n}\right]
\textstyle{\frac{\mp1}{{2\;}}}\left[
{\bf n}({\bf e}_x\pm i{\bf e}_y)
+({\bf e}_x\pm i{\bf e}_y){\bf n}
\right]
\\
{\bf t}_{0,0}&\equiv&\textstyle{\frac{1}{\sqrt{3\;}}}
\left[{\bf e}_x{\bf e}_x+{\bf e}_y{\bf e}_y+{\bf e}_z{\bf e}_z\right]\\
{\bf t}_{2,0}&\equiv&\textstyle{\frac{-1}{\sqrt{6\;}}}
\left[{\bf e}_x{\bf e}_x+{\bf e}_y{\bf e}_y-2{\bf e}_z{\bf e}_z\right]\\
{\bf t}_{2,\pm1}&\equiv&\mp\textstyle{\frac{1}{2}}
\left[{\bf e}_x{\bf e}_z+{\bf e}_z{\bf e}_x\right]
-\textstyle{\frac{1}{2}}\,i
\left[{\bf e}_y{\bf e}_z+{\bf e}_z{\bf e}_y\right]\\
{\bf t}_{2,\pm2}&\equiv&\textstyle{\frac{1}{2}}
\left[{\bf e}_x{\bf e}_x-{\bf e}_y{\bf e}_y\pm i\left({\bf e}_y{\bf e}_x+{\bf e}_x{\bf e}_y
\right)
\right]\,.\label{t22def}
\end{eqnarray}

We next define the corotating equivalents to the basis vectors
\begin{equation}
\widetilde{\bf n}={\bf n}\quad
\widetilde{\bf e}_x={\bf e}_x\cos{\Omega t}+{\bf e}_y\sin{\Omega t}\quad
\widetilde{\bf e}_x=-{\bf e}_x\sin{\Omega t}+{\bf e}_y\cos{\Omega t}\quad
\widetilde{\bf e}_z={\bf e}_z, 
\end{equation}
and we use these to define the corotating equivalents of the basis tensors:
\begin{eqnarray}
\widetilde{\bf t}_{nn}&\equiv&\widetilde{\bf n}\widetilde{\bf n}
={\bf t}_{nn}\\
\widetilde{\bf t}_{n0}&\equiv&\widetilde{\bf n}\widetilde{\bf e}_z
={\bf t}_{n0}\\
\widetilde{\bf t}_{n,\pm1}&\equiv&
%\textstyle{\frac{1}{\sqrt{2\;}}}\left[\widetilde{\bf n}\widetilde{\bf s}_{\pm1}
%+\widetilde{\bf s}_{\pm1}\widetilde{\bf n}\right]
\textstyle{\frac{\mp1}{{2\;}}}\left[
\widetilde{\bf n}(\widetilde{\bf e}_x\pm i\widetilde{\bf e}_y)
+(\widetilde{\bf e}_x\pm i\widetilde{\bf e}_y)\widetilde{\bf n}\right]
=e^{\mp i\Omega t}
{\bf t}_{n,\pm1}
\\
\widetilde{\bf t}_{0,0}&\equiv&\textstyle{\frac{1}{\sqrt{3\;}}}
\left[\widetilde{\bf e}_x\widetilde{\bf e}_x+\widetilde{\bf e}_y\widetilde{\bf e}_y
+\widetilde{\bf e}_z\widetilde{\bf e}_z\right]={\bf t}_{0,0}\\
\widetilde{\bf t}_{2,0}&\equiv&\textstyle{\frac{-1}{\sqrt{6\;}}}
\left[\widetilde{\bf e}_x\widetilde{\bf e}_x+\widetilde{\bf e}_y\widetilde{\bf e}_y
-2\widetilde{\bf e}_z\widetilde{\bf e}_z\right]={\bf t}_{2,0}\\
\widetilde{\bf t}_{2,\pm1}&
\equiv&\mp\textstyle{\frac{1}{2}}
\left[\widetilde{\bf e}_x\widetilde{\bf e}_z
+\widetilde{\bf e}_z\widetilde{\bf e}_x\right]
-\textstyle{\frac{1}{2}}\,i
\left[\widetilde{\bf e}_y\widetilde{\bf e}_z+\widetilde{\bf e}_z
\widetilde{\bf e}_y\right]
=e^{\mp i\Omega t}
{\bf t}_{2,\pm1}
\\
\widetilde{\bf t}_{2,\pm2}&\equiv&\textstyle{\frac{1}{2}}
\left[\widetilde{\bf e}_x\widetilde{\bf e}_x-\widetilde{\bf e}_y
\widetilde{\bf e}_y
\pm i\left(\widetilde{\bf e}_y\widetilde{\bf e}_x+\widetilde{\bf e}_x
\widetilde{\bf e}_y
\right)
\right]=
e^{\mp2i\Omega t}
{\bf t}_{2,\pm2}
\,.
\end{eqnarray}
Note that  all the tensor bases are orthonormal in the following sense:
\begin{equation}\label{orthonorm} 
({\bf t}_{i,j})^*\cdot({\bf t}_{k,p})=(\widetilde{\bf t}_{i,j})^*
\cdot(\widetilde{\bf t}_{k,p})
=\delta_{ik}\delta_{jp}\,,
\end{equation}
and that the tensor $\bar{h}
$ can be written as 
\begin{displaymath}
{\bf \bar{h}}=\Psi^{(nn)}{\bf t}_{nn}+\Psi^{(n0)}{\bf t}_{n0}
+\Psi^{(n1)}{\bf t}_{n1}
+\Psi^{(n,-1)}{\bf t}_{n,-1}+\Psi^{(00)}{\bf t}_{00}+\Psi^{(20)}{\bf t}_{20}+
\end{displaymath}
\begin{equation}\label{hbarinertial} 
\Psi^{(21)}{\bf t}_{21}
+\Psi^{(2,-1)}{\bf t}_{2,-1}
+\Psi^{(22)}{\bf t}_{22}+\Psi^{(2,-2)}{\bf t}_{2,-2}\,,
\end{equation}
or as
\begin{displaymath}
{{\bf \bar{h}}}= \widetilde{\Psi}^{(nn)}\widetilde{\bf
t}_{nn}+\widetilde{\Psi}^{(n0)}\widetilde{\bf t}_{n0}
+\widetilde{\Psi}^{(n1)}\widetilde{\bf t}_{n10}
+\widetilde{\Psi}^{(n,-10)}\widetilde{\bf t}_{n,-1}
+\widetilde{\Psi}^{(00)}\widetilde{\bf t}_{00}
+\widetilde{\Psi}^{(20)}\widetilde{\bf t}_{20}+
\end{displaymath}
\begin{equation}\label{hbartilde} 
\widetilde{\Psi}^{(21)}\widetilde{\bf t}_{21}
+\widetilde{\Psi}^{(2,-1)}\widetilde{\bf t}_{2,-1}
+\widetilde{\Psi}^{(22)}\widetilde{\bf t}_{22}
+\widetilde{\Psi}^{(2,-2)}\widetilde{\bf t}_{2,-2}\ .
\end{equation}

From Eq.~(\ref{orthonorm}) we get
\begin{eqnarray}
\Psi^{(nn)}&=&\bar{h}^{tt}\label{Psifirst}\\
\Psi^{(n0)}&=&\bar{h}^{tz}\\
\Psi^{(n1)}&=&-\bar{h}^{tx}+i\bar{h}^{ty}\\
\Psi^{(00)}&=&\textstyle{\frac{1}{\sqrt{3\;}}}\left[
\bar{h}^{xx}+\bar{h}^{yy}+\bar{h}^{zz}
\right]\\
\Psi^{(20)}&=&\textstyle{\frac{-1}{\sqrt{6\;}}}\left[
\bar{h}^{xx}+\bar{h}^{yy}-2\bar{h}^{zz}
\right]\\
\Psi^{(21)}&=&-
\bar{h}^{xz}
+i\bar{h}^{yz}\\
\Psi^{(22)}&=&\textstyle{\frac{1}{2}}\left[
\bar{h}^{xx}-\bar{h}^{yy}\right]
-i\bar{h}^{yx}\label{Psilast}\ .
\end{eqnarray}
Here and below, we have dropped functions that are 
redundant due to the relations
\begin{equation}
\Psi^{(n,-1)}=-\,\left(\Psi^{(n1)}\right)^*
\quad\quad\quad
\Psi^{(2,-1)}=-\,\left(\Psi^{(21)}\right)^*
\quad\quad\quad
\Psi^{(2,-2)}=\left(\Psi^{(22)}\right)^*\,.
\end{equation}
These relations are also true if tildes are placed over all variables.
From the rotation behaviors of the $\widetilde{\bf t} $ basis, given
above, the rotation laws for the $\Psi$ fields are:
\begin{eqnarray}
\widetilde\Psi^{(nn)}&=&{\Psi}^{(nn)}\label{tildePsifirst}\\
\widetilde\Psi^{(n0)}&=&{\Psi}^{(n0)}\\
\widetilde\Psi^{(n1)}&=&e^{i\Omega t}\;{\Psi}^{(n1)}=U^{(n1)}+iV^{(n1)}  \\
\widetilde\Psi^{(00)}&=&{\Psi}^{(00)}\\
\widetilde\Psi^{(20)}&=&{\Psi}^{(20)}\label{tildePsi20}\\
\widetilde\Psi^{(21)}&=&e^{i\Omega t}\;{\Psi}^{(21)}=U^{(21)}+iV^{(21)}
\label{tildePsipenult}        \\
\widetilde\Psi^{(22)}&=&e^{2i\Omega t}\;{\Psi}^{(22)}=U^{(22)}+iV^{(22)}\,,
\label{tildePsilast}
\end{eqnarray}
where $U^{(ab)}$ and $V^{(ab)}$ are real functions. 

We define projections of the stress energy by analogy with the
projections of the $\bar{h} $ perturbations,
\begin{eqnarray}
{\cal T}^{(nn)}&=&{T}^{tt}=
\widetilde{\cal T}^{(nn)}\label{Tfirst}  \\
{\cal T}^{(n0)}&=&{T}^{tz}=
\widetilde{\cal T}^{(n0)}\\
{\cal T}^{(n1)}&=&-{T}^{tx}+i\,{T}^{ty}
=e^{-i\Omega t}
\widetilde{\cal T}^{(n1)}\\
{\cal T}^{(00)}&=&\textstyle{\frac{1}{\sqrt{3\;}}}\left[
{T}^{xx}+{T}^{yy}+{T}^{zz}
\right]=
\widetilde{\cal T}^{(00)}\\
{\cal T}^{(20)}&=&\textstyle{\frac{-1}{\sqrt{6\;}}}\left[
{T}^{xx}+{T}^{yy}-2{T}^{zz}
\right]=
\widetilde{\cal T}^{(20)}\\
{\cal T}^{(21)}&=&-
{T}^{xz}
+i\,{T}^{yz}
=e^{-i\Omega t}
\widetilde{\cal T}^{(21)}\\
{\cal T}^{(22)}&=&\textstyle{\frac{1}{2}}\left[
{T}^{xx}-{T}^{yy}\right]
-i\,{T}^{yx}
=e^{-2i\Omega t}
\widetilde{\cal T}^{(22)}\label{Tlast}
\ .
\end{eqnarray}

Since the inertial basis tensors are covariantly constant, we can 
write the field equations (\ref{boxhbareqT}) as 
\begin{equation}
\Box \Psi^{(ab)}=-16\pi {\cal T}^{(ab)}
\end{equation}
where $\Box$ is the simple scalar d'Alembertian. From the relations in
Eqs.~(\ref{tildePsifirst})--(\ref{tildePsilast}) and
(\ref{Tfirst})--(\ref{Tlast}) we can then write
\begin{equation}
e^{ik\Omega t}
\Box\left(e^{-ik\Omega t}\widetilde{\Psi}^{(ab)}\right)=
-16\pi\widetilde{\cal T}^{(ab)}\,.
\end{equation}
with $k=0\pm1,\pm2$.

Lastly, for a helical scalar $f$
we have that 
\begin{equation}
e^{ik\Omega t}\Box\left(e^{-ik\Omega t}f\right)
=\Box f+2ik\Omega\partial_tf+k^2\Omega^2f\,.
\end{equation}
The time derivative $\partial_tf$ here uses the time coordinate of the
inertial $t,r,\theta,\phi $ system and, for $f $ a (helically
symmetric) function of the corotating coordinates $t,r,\theta,\varphi
$, is equivalent to $-\Omega\partial_\varphi f$. The field equations
for the metric perturbations then take the form
\begin{eqnarray}
\Box\widetilde{\Psi}^{(ab)}-2ik\Omega^2\partial_\varphi\widetilde{\Psi}
+k^2\Omega^2\widetilde{\Psi}=-16\pi\widetilde{\cal T}^{(ab)}
\end{eqnarray}
where $\Box$ is the scalar d'Alembertian.
The explicit equations are

\begin{eqnarray}
\Box\widetilde{\Psi}^{(00)}
&=&-16\pi\widetilde{\cal T}^{(00)}\label{BoxWT00} \\
\Box\widetilde{\Psi}^{(n0)}&=&-16\pi\widetilde{\cal T}^{(n0)}\\
\Box\widetilde{\Psi}^{(20)}&=&-16\pi\widetilde{\cal T}^{(20)}\\
\Box\widetilde{\Psi}^{(nn)}&=&
-16\pi\widetilde{\cal T}^{(nn)}\\
\Box\widetilde{\Psi}^{(n1)}
-2i\Omega^2\partial_{\varphi}\widetilde{\Psi}^{(n1)}
+\Omega^2\widetilde{\Psi}^{(n1)}&=&-16\pi\widetilde{\cal T}^{(n1)}\\
\Box\widetilde{\Psi}^{(21)}
-2i\Omega^2\partial_{\varphi}\widetilde{\Psi}^{(21)}
+\Omega^2\widetilde{\Psi}^{(21)}&=&-16\pi\widetilde{\cal T}^{(21)}\\
\Box\widetilde{\Psi}^{(22)}
-4i\Omega^2\partial_{\varphi}\widetilde{\Psi}^{(22)}
+4\Omega^2\widetilde{\Psi}^{(22)}&=&
-16\pi\widetilde{\cal T}^{(22)}\,.\label{BoxWT22} 
\end{eqnarray}

From Eqs.~(\ref{barhinnerfirst})--  (\ref{barhinnerlast}),
and the prescriptions in 
Eqs.~(\ref{Psifirst})--  (\ref{Psilast}),
and Eqs.~(\ref{tildePsifirst})--  (\ref{tildePsilast}),
we have the inner boundary conditions, to be applied 
at small $\chi$ to represent the near field of the  mass points
at $\chi=0
$\,,
\begin{eqnarray}
\widetilde{\Psi}^{(nn)}&=&4m_0
\frac{\gamma^2}
{\sqrt{\widetilde{r}^2+\gamma^2v^2\widetilde{y}^2\;}}
\label{innerfirst}\\
\widetilde{\Psi}^{(00)}&=&4m_0\frac{\gamma^2
}{\sqrt{\widetilde{r}^2+\gamma^2v^2\widetilde{y}^2\;}}
\;\frac{v^2
}{\sqrt{3\;}}
\\
\widetilde{\Psi}^{(20)}&=&4m_0\frac{\gamma^2
}{\sqrt{\widetilde{r}^2+\gamma^2v^2\widetilde{y}^2\;}}
\left(\,-\,\frac{v^2
}{\sqrt{6\;}}\right)
\\
V^{(n1)}&=&4m_0\frac{\gamma^2
}{\sqrt{\widetilde{r}^2+\gamma^2v^2\widetilde{y}^2\;}}
\;v\; {\rm sgn}[\cos{\Theta}]
\\
%
%\widetilde{\Psi}^{(21)}&=&0\\
{U}^{(22)}&=&4m_0\frac{\gamma^2
}{\sqrt{\widetilde{r}^2+\gamma^2v^2\widetilde{y}^2\;}}
\left(\,-\,\frac{v^2
}{2}
\right)
\end{eqnarray}
\begin{equation}\label{innerlast}
\widetilde{\Psi}^{(n0)}
=U^{(21)}=V^{(21)}=
U^{(n1)}=V^{(22)}=0\,.
\end{equation}

\subsection{Outer boundary conditions for helical scalars}

The quantities $\Psi^{(nn)}$, $\Psi^{(00)}$, $\Psi^{(n0)}$, $\Psi^{(20)}$, 
are nonradiative multipoles that fall off as 
$1/r$. The radiative parts of these fields satisfy a Sommerfeld 
condition
\begin{equation}\label{radpartnn} 
\frac{\partial}{\partial r}\left[r
\widetilde{\Psi}\right]
=\pm\Omega\frac{\partial}{\partial\varphi}\left[r\widetilde{\Psi}\right]\,,
\end{equation}
in which the upper and lower signs correspond respectively to outgoing
and ingoing waves.  In practice, special attention is not necessary
for the nonradiatable multipoles of $\Psi^{(nn)}$, $\Psi^{(00)}$,
$\Psi^{(n0)}$, $\Psi^{(20)}$. The Sommerfeld conditions can be applied
to the total field, without regard to multipole content.

Some care must be taken with the Sommerfeld conditions for
$\widetilde{\Psi}^{(n1)}$, $\widetilde{\Psi}^{(21)}$ and
$\widetilde{\Psi}^{(22)}$.  The Sommerfeld condition applies to the
inertial projections of $\bar{h}_{\alpha\beta}$, and hence to
${\Psi}^{(n1)}$, ${\Psi}^{(21)}$ and ${\Psi}^{(22)}$. With the
relations, in Eqs.~(\ref{tildePsifirst})--
(\ref{tildePsilast}), between these quantities and the ``helical
scalars'' $\widetilde{\Psi}^{(n1)}$, $\widetilde{\Psi}^{(21)}$,
$\widetilde{\Psi}^{(22)}$ used in computation, we arrive at the
conditions
\begin{equation}\label{outhelfirst}
\frac{\partial}{\partial r}\left[re^{i\varphi}
\widetilde{\Psi}^{(n1)}\right]
=\pm\Omega\frac{\partial}{\partial\varphi}
\left[re^{i\varphi}\widetilde{\Psi}^{(n1)}\right]
\end{equation}
\begin{equation}
\frac{\partial}{\partial r}\left[re^{i\varphi}
\widetilde{\Psi}^{(21)}\right]
=\pm\Omega\frac{\partial}{\partial\varphi}
\left[re^{i\varphi}\widetilde{\Psi}^{(21)}\right]
\end{equation}
\begin{equation}\label{outhellast}
\frac{\partial}{\partial r}\left[re^{2i\varphi}
\widetilde{\Psi}^{(22)}\right]
=\pm\Omega\frac{\partial}{\partial\varphi}
\left[re^{2i\varphi}\widetilde{\Psi}^{(22)}\right]\,,
\end{equation}
or
\begin{eqnarray}
\frac{1}{r}\,\frac{\partial}{\partial r}\left(rU^{(ab)}
\right)=\pm\Omega\left(
-kV^{(ab)}+\frac{\partial U^{(ab)}}{\partial\varphi}
\right)\label{outbcU}  \\
\frac{1}{r}\,\frac{\partial}{\partial r}\left(rV^{(ab)}
\right)=\pm\Omega\left(
kU^{(ab)}+\frac{\partial V^{(ab)}}{\partial\varphi}
\right)\label{outbcV}\,,
\end{eqnarray}
in which $k=1$ for $(ab)=(n1)$ or $(21)$, and $k=2$ for $(ab)=(22)$.
\subsection{Series solutions}\label{sub:series}

To develop series solutions of Eqs.~(\ref{BoxWT00}) -- (\ref{BoxWT22})
we start by using the expressions in Eqs.~(\ref{T00value}) --
(\ref{Tijvalue}), for stress-energy components of the symmetric pair
of particles, in the general expressions in Eqs.~(\ref{Tfirst}) --
(\ref{Tlast}). The nonvanishing results are
\begin{eqnarray}
\widetilde{\cal T}^{(nn)}
&=&m_0\,
\gamma
\;\frac{\delta\left(r-a\right)}{a^2}
\delta\left(\theta-\pi/2 \right)
\left[
\delta\left(\varphi\right)+\delta\left(\varphi-\pi\right)
\right]\\
\widetilde{\cal T}^{(n1)}&=&\pm i\,v\widetilde{\cal T}^{(nn)}\\
\widetilde{\cal T}^{(20)}=- \frac{v^2}{\sqrt{6\;}}\widetilde{\cal T}^{(nn)}
&\quad\quad& \widetilde{\cal T}^{(00)}=
\frac{v^2}{\sqrt{3\;}}\widetilde{\cal T}^{(nn)}
\\
\widetilde{\cal T}^{(22)}&=& -\,\frac{v^2}{2}\widetilde{\cal T}^{(nn)}\ .
\end{eqnarray}
As in Eq.~(\ref{T0jval}), the upper sign in the expression for
$\widetilde{T}^{(n1)}$ indicates the particle at $\varphi=0$, the 
lower sign indicates the particle at $\varphi=\pi$.

With these expressions, Eqs.~(\ref{BoxWT00}) -- (\ref{BoxWT22})
take the explicit form
\begin{equation}\label{Boxnn} 
\Box\widetilde{\Psi}^{(nn)}=
-16\pi m_0\frac{\delta(r-a)}{a^2}\delta(\cos\theta)
\,\gamma\,\left[
\delta(\varphi)
+
\delta(\varphi-\pi)
\right]
\end{equation}
\begin{equation}
\widetilde{\Psi}^{(n0)}=0
\end{equation}
\begin{equation}
\Box\widetilde{\Psi}^{(00)}=
-16\pi m_0\frac{\delta(r-a)}{a^2}\delta(\cos\theta)
\,\frac{v^2\gamma}{\sqrt{3\;}}\,\left[
\delta(\varphi)
+
\delta(\varphi-\pi)
\right]
\end{equation}
\begin{equation}\label{Box20} 
\Box\widetilde{\Psi}^{(20)}=
+\,16\pi m_0\frac{\delta(r-a)}{a^2}\delta(\cos\theta)
\,\frac{v^2\gamma}{\sqrt{6\;}}\,\left[
\delta(\varphi)
+
\delta(\varphi-\pi)
\right]
\end{equation}

\smallskip
\begin{equation}
\Box\widetilde{\Psi}^{(n1)}-2i\Omega^2\partial_{\varphi}\widetilde{\Psi}^{(n1)}
+\Omega^2\widetilde{\Psi}^{(n1)}=
-\,i\,16\pi m_0\frac{\delta(r-a)}{a^2}\delta(\cos\theta)
\,{v\gamma}
\,\left[
\delta(\varphi)
-
\delta(\varphi-\pi)
\right]
\end{equation}

\begin{equation}
\widetilde{\Psi}^{(21)}=0
\end{equation}
\begin{equation}\label{Box22} 
\Box\widetilde{\Psi}^{(22)}
-4i\Omega^2\partial_{\varphi}\widetilde{\Psi}^{(22)}
+4\Omega^2\widetilde{\Psi}^{(22)}=
+\,16\pi m_0\frac{\delta(r-a)}{a^2}\delta(\cos\theta)
\,\frac{v^2
\gamma}{2}
\,\left[
\delta(\varphi)
+
\delta(\varphi-\pi)
\right]
\end{equation}
where
\begin{equation}
\Box\widetilde{\Psi}=
\frac{1}{r^2}\frac{\partial}{\partial r}
\left(r^2\frac{\partial\widetilde{\Psi}}{\partial r}
\right)+\frac{1}{r^2\sin\theta}\frac{\partial}{\partial\theta}
\left(\sin\theta\frac{\partial\widetilde{\Psi}}{\partial\theta}
\right)+\left(\frac{1}{r^2\sin^2\theta
}-\Omega^2
\right)\frac{\partial^2\widetilde{\Psi}
}{\partial\varphi^2}\ .
\end{equation}

For $\widetilde{\Psi}^{(nn)}$, 
$\widetilde{\Psi}^{(20)}$, $\widetilde{\Psi}^{(00)}$
the equations all have the form
\begin{equation}\label{Kdef} 
\Box\widetilde{\Psi}=K\frac{\delta(r-a)}{a^2
}\delta(\cos\theta)\left[\delta(\varphi)+\delta(\varphi-\pi)
\right]
\end{equation}
in which the value of $K$ can be read from Eqs.~(\ref{Boxnn})
-- (\ref{Box20}).
The outoging solutions are constructed in 
the usual manner from the spherical Bessel and Hankel functions:
\begin{equation}\label{allPsicomplex} 
\Psi=-2i\Omega K\sum_{\ell,m\;\rm even }mj_{\ell}(m\Omega r_<)
h^{(1)}_{\ell}(m\Omega r_>)Y^*_{\ell m}(\pi/2,0)Y_{\ell m}(\theta,\varphi)
\end{equation}
or
\begin{displaymath}
\Psi=-2K\,
 \sum_\ell \frac{1}{2\ell+1}\;Y_{\ell 0}^*(\pi/2,0)Y_{\ell 0}(\theta,0)
\;\frac{r_<^\ell
}{r^{\ell+1}_>}
\hspace{2in}
\end{displaymath}
\begin{equation}\label{allPsi} 
+4K\Omega\sum_\ell 
\sum_{m={\rm 2,4,6}}m\;Y_{\ell m}^*(\pi/2,0)Y_{\ell m}(\theta,0)
j_\ell(m\Omega r_<)
{\rm Im}\left\{h^{(1)}_\ell(m\Omega r_>)e^{im\varphi}\right\}\,.
\end{equation}

By expansion in spherical harmonics, then by the usual Green function
construction, the outgoing solution for $\widetilde{\Psi}^{(n1)}
$ is found to be
\begin{displaymath}
\widetilde{\Psi}^{(n1)}=U^{(n1)}+iV^{(n1)}=
-\,32\pi m_0v\gamma\;
\sum_{\ell}\left[
Y_{\ell, -1}^*(\pi/2,0)Y_{\ell, -1}(\theta,0)
\left(\frac{-ie^{-i\varphi}}{2\ell+1}\right)\;\frac{r_<^{\ell}}{r_>^{\ell+1}}
\right.
\end{displaymath}
\begin{equation}\label{Psin1} 
\left.+ \Omega\sum_{m={\rm odd},\neq-1} (m+1)\;Y_{\ell, m}^*(\pi/2,0)
Y_{\ell, m}(\theta,0)j_\ell((m+1)\Omega r_<)h^{(1)}_\ell((m+1)\Omega r_>)
e^{im\varphi}\right]
\end{equation}
%
%%%%%%%%%%%%%%%%%%%%%%%%%%%%%%%%%%%%%%%%%%%
or
\begin{displaymath}
U^{(n1)}=-32\pi m_0v\gamma\sum_{\ell\;\rm odd}\Biggl[
\frac{-\sin{\varphi}}{(2\ell+1)}\frac{r_<^\ell}{r_>^{\ell+1}
}\;Y^*_{\ell -1}(\pi/2,0)Y_{\ell -1}(\theta,0)
\Biggr.
\end{displaymath}
\begin{equation}\label{Un1soln} 
+\Omega\!\!\!\!\!\sum_{m=1,3,5,\cdots}
\!\!\!\!\!(m\!+\!1) Y^*_{\ell m}(\pi/2,0)Y_{\ell m}(\theta,0)
j_\ell((m\!+\!1)\Omega r_<)[\cos{m\varphi}\,j_\ell((m\!+\!1)\Omega r_>)
-\sin{m\varphi}\,n_\ell((m\!+\!1)\Omega r_>)]
\end{equation}
\begin{displaymath}
\left.-\Omega\!\!\!\!\!\sum_{m=1,3,5,\cdots}
\!\!\!\!\!(m\!+\!1) Y^*_{\ell,\!-m\!-\!2}(\pi/2,0)Y_{\ell,\!-m\!-\!2}(\theta,0)
j_\ell((m\!+\!1)\Omega r_<)[\cos{((m\!+\!2)\varphi)}\,j_\ell((m\!+\!1)\Omega r_>)
-\sin{((m\!+\!2)\varphi)}\,n_\ell((m\!+\!1)\Omega r_>)]\right]
\end{displaymath}
%
%
%%%%%%%%%%%%%%%%%%%%%%%%%%%%%%%%%%
and
\begin{displaymath}
V^{(n1)}=-32\pi m_0v\gamma\sum_{\ell\;\rm odd}\Biggl[
\frac{-\cos{\varphi}}{(2\ell+1)}\frac{r_<^\ell}{r_>^{\ell+1}
}\;Y^*_{\ell, -1}(\pi/2,0)Y_{\ell, -1}(\theta,0)
\Biggr.
\end{displaymath}
\begin{equation}\label{Vn1soln} 
+\Omega\!\!\!\!\!\sum_{m=1,3,5\cdots}\!\!\!\!\!
(m\!+\!1) Y^*_{\ell m}(\pi/2,0)Y_{\ell m}(\theta,0)
j_\ell((m\!+\!1)\Omega r_<)[\cos{m\varphi}\,n_\ell((m\!+\!1)\Omega r_>)
+\sin{m\varphi}\,j_\ell((m\!+\!1)\Omega r_>)]
\end{equation}
\begin{displaymath}
\left.+\Omega\!\!\!\!\!\sum_{m=1,3,5\cdots}\!\!\!\!\!
(m\!+\!1) Y^*_{\ell,\!-\!m\!-\!2}(\pi/2,0)Y_{\ell,\!-\!m\!-\!2}(\theta,0)
j_\ell((m\!+\!1)\Omega r_<)[\cos{((m\!+\!2)\varphi)}\,n_\ell((m\!+\!1)\Omega r_>)
+\sin{((m\!+\!2)\varphi)}\,j_\ell((m\!+\!1)\Omega r_>)]\right]\,.
\end{displaymath}
%
%%%%%%%%%%%%%%%%%%%%%%%%%%%%%%%%%%
Similarly, the outgoing solution for $\widetilde{\Psi}^{(22)}
$ is found to be
\begin{displaymath}
\widetilde{\Psi}^{(22)}=U^{(22)}+iV^{(22)}=
\,16\pi m_0v^2
\gamma\;
\sum_{\ell=\rm even}\left[
Y_{\ell,-2}^*(\pi/2,0)Y_{\ell,-2}(\theta,0)
\left(\frac{-e^{-2i\varphi}}{2\ell+1}\right)\;
\frac{r_<^\ell
}{r_>^{\ell+1}}\right.
\end{displaymath}
\begin{equation}\label{Psi22soln} 
\left.-i \Omega\!\!\!\!\!\!
\sum_{m={\rm even},\neq-2} \!\!\!\!\!
(m\!+\!2)\;Y_{\ell, m}^*(\pi/2,0)Y_{\ell, m}(\theta,0)
j_\ell((m\!+\!2)\Omega r_<)h^{(1)}_\ell((m\!+\!2)\Omega r_>)
e^{im\varphi}\right]
\end{equation}
or
\begin{displaymath}
U^{(22)}=16\pi m_0v^2
\gamma\sum_{\ell\;\rm even}\Biggl[
\frac{-\cos{2\varphi}}{(2\ell+1)}\frac{r_<^\ell}{r_>^{\ell+1}
}\;Y^*_{\ell,-2}(\pi/2,0)Y_{\ell,-2}(\theta,0)
\Biggr.
\end{displaymath}
\begin{equation}\label{U22soln} 
+\Omega\!\!\!\!\!\!
\sum_{m=0,2,4\cdots}\!\!\!\!\!
(m\!+\!2) Y^*_{\ell m}(\pi/2,0)Y_{\ell m}(\theta,0)
j_\ell((m\!+\!2)\Omega r_<)[\cos{m\varphi}\,n_\ell((m\!+\!2)\Omega r_>)
+\sin{m\varphi}\,j_\ell((m\!+\!2)\Omega r_>
)]
\end{equation}
\begin{displaymath}
\left.+\Omega\!\!\!\!\!\!
\sum_{m=0,2,4\cdots}\!\!\!\!\!
(m\!+\!2) Y^*_{\ell,-m-4}(\pi/2,0)Y_{\ell,-m-4}(\theta,0)
j_\ell((m\!+\!2)\Omega r_<)[\cos{((m\!+\!4)\varphi)}\,n_\ell((m\!+\!2)\Omega r_>)
+\sin{((m\!+\!4)\varphi)}\,j_\ell((m\!+\!2)\Omega r_>
)]\right]
\end{displaymath}
and
\begin{displaymath}
V^{(22)}=16\pi m_0v^2
\gamma\sum_{\ell\;\rm even}\Biggl[
\frac{\sin{2\varphi}}{(2\ell+1)}\frac{r_<^\ell}{r_>^{\ell+1}
}\;Y^*_{\ell,-2}(\pi/2,0)Y_{\ell,-2}(\theta,0)
\Biggr.
\end{displaymath}
\begin{equation}\label{V22soln} 
+\Omega\!\!\!\!\!\!
\sum_{m=0,2,4\cdots}\!\!\!\!\!\!
(m\!+\!2) Y^*_{\ell m}(\pi/2,0)Y_{\ell m}(\theta,0)
j_\ell((m\!+\!2)\Omega r_<)
[
\sin{m\varphi}\,n_\ell((m\!+\!2)\Omega r_>)
-\cos{m\varphi}\,j_\ell((m\!+\!2)\Omega r_>)]
\end{equation}
\begin{displaymath}
\left.-\Omega\!\!\!\!\!\!
\sum_{m=0,2,4\cdots}\!\!\!\!\!\!
(m\!+\!2) Y^*_{\ell,\!-m\!-\!4}(\pi/2,0)Y_{\ell,-\!m-\!4}(\theta,0)
j_\ell((m\!+\!2)\Omega r_<)
[
\sin{((m\!+\!4)\varphi)}\,n_\ell((m\!+\!2)\Omega r_>)
-\cos{((m\!+\!4)\varphi)}\,j_\ell((m\!+\!2)\Omega r_>)]
\right]\,.
\end{displaymath}
The expressions for $U^{(n1)}$, $V^{(n1)}$, $U^{(22)}$, and $V^{(22)}$
have been given as sums only over nonnegative values of $m
$ through the use of the relationship $j_\ell(-x)h_\ell^{(1)}(-x)
=j_\ell(x)(h_\ell^{(1)}(x))^*$.

\section{Linearized Gravity in Adapted
Coordinates}\label{sec:adapcoord} \subsection{Field equations in
adapted coordinates}
The field equations to be used for computation are the source-free forms of 
Eqs.~(\ref{BoxWT00})
--
(\ref{BoxWT22}).
The first four of  these equations are
\begin{eqnarray}\label{BoxWTPsi00} 
\Box\widetilde{\Psi}^{(ab)}&=&0
\end{eqnarray}
for $(ab)
$=$(00)$, $(n0)$, $(20)$, and $(n0)$.
In real form the last three of these equations are
\begin{eqnarray}
\Box{U}^{(ab)}
+2k\Omega^2\partial_{\varphi}{V}^{(ab)}
+k^2\Omega^2{U}^{(ab)}&=&0\label{BoxU}  \\
\Box{V}^{(ab)}
-2k\Omega^2\partial_{\varphi}{U}^{(ab)}
+k^2\Omega^2{V}^{(ab)}&=&0\label{BoxV} 
\end{eqnarray}
with $k=1 $ for $(ab) $=$(n1)$, or $(21)$, and $k=2$ for $(ab)=(22)$.

The form of the scalar d'Alembertian in adapted coordinates is given in
Eq.~(8) of
Ref.~\cite{eigenspec}, as
\begin{displaymath}
\Box\Psi=A_{\chi\chi}\;\frac{\partial^2\Psi}{\partial\chi^2}
+A_{\Theta\Theta}\;\frac{\partial^2\Psi}{\partial\Theta^2}
+A_{\Phi\Phi}\;\frac{\partial^2\Psi}{\partial\Phi^2}
+2A_{\chi\Theta}\;\frac{\partial^2\Psi}{\partial\chi\partial\Theta}
+2A_{\chi\Phi}\;\frac{\partial^2\Psi}{\partial\chi\partial\Phi}
+2A_{\Theta\Phi}\;\frac{\partial^2\Psi}{\partial\Theta\partial\Phi}
\end{displaymath}
\begin{equation}\label{waveq} 
+B_{\chi}\;\frac{\partial\Psi}{\partial\chi}
+B_{\Theta}\;\frac{\partial\Psi}{\partial\Theta}
+B_{\Phi}\;\frac{\partial\Psi}{\partial\Phi}\,,
\end{equation}
and the adapted-coordinate form of $\partial _\varphi$
in Eq.~(27) of Ref.~\cite{eigenspec},
\begin{equation}\label{FDMoutbc} 
\frac{\partial}{\partial\varphi}=
\left(\Gamma^\Theta\frac{\partial\Psi}{\partial\Theta}
+\Gamma^\Phi\frac{\partial\Psi}{\partial\Phi}
+\Gamma^\chi\frac{\partial\Psi}{\partial\chi} \right)\,.
\end{equation}
Here the $A$, $B$ and $\Gamma
$ coefficients are known, real functions of
$\chi,\Theta,\Phi$ that are given explicitly in Appendix A of
Ref.~\cite{eigenspec} and are repeated in the appendix of the 
present paper.

\subsection{Inner boundary conditions  in adapted coordinates}

To express the inner boundary conditions in terms of the adapted 
coordinates, we approximate
\begin{displaymath}
\widetilde{r}^2+\gamma^2v^2\widetilde{y}^2
=(\widetilde{Z}-a)^2+\widetilde{X}^2+\widetilde{Y}^2
+\gamma^2v^2\widetilde{X}^2=\hspace{1in}
\end{displaymath}
\begin{equation}
\hspace{1in}
\left[1+\gamma^2v^2\sin^2{2\Theta}\cos^2\Phi\right]\,
\frac{\chi^4}{4a^2}+{\cal O}(\chi^6/a^4)\,,
\end{equation}
and we write Eqs.~(\ref{innerfirst})--(\ref{innerlast}), for the $\chi\rightarrow0
$ limits of the the fields, as
\begin{eqnarray}
\widetilde{\Psi}^{(nn)}&=&4m_0
\frac{2a\gamma^2}
{\chi^2\sqrt{1+\gamma^2v^2\sin^2{2\Theta}\cos^2\Phi\;}}
\label{innerfirstadap}\\
\widetilde{\Psi}^{(00)}&=&4m_0\frac{2a\gamma^2
}{\chi^2\sqrt{1+\gamma^2v^2\sin^2{2\Theta}\cos^2\Phi\;}}
\;\frac{v^2
}{\sqrt{3\;}}
\\
\widetilde{\Psi}^{(20)}&=&4m_0\frac{2a\gamma^2
}{\chi^2\sqrt{1+\gamma^2v^2\sin^2{2\Theta}\cos^2\Phi\;}}
\left(\,-\,\frac{v^2
}{\sqrt{6\;}}\right)\label{inner20adap}
\\
V^{(n1)}&=&4m_0\frac{2a\gamma^2
}{\chi^2\sqrt{1+\gamma^2v^2\sin^2{2\Theta}\cos^2\Phi\;}}
\;v\; {\rm sgn}[\cos{\Theta}]
\\
%
%\widetilde{\Psi}^{(21)}&=&0\\
{U}^{(22)}&=&4m_0\frac{2a\gamma^2
}{\chi^2\sqrt{1+\gamma^2v^2\sin^2{2\Theta}\cos^2\Phi\;}}
\left(\,-\,\frac{v^2
}{2}
\right)
\end{eqnarray}
\begin{equation}\label{innerlastadap}
\widetilde{\Psi}^{(n0)}
=U^{(21)}=V^{(21)}=
U^{(n1)}=V^{(22)}=0\,.
\end{equation}

\subsection{Outer boundary conditions in adapted coordinates}

The outer boundary conditions in adapted coordinates follow from
Eqs.~(\ref{radpartnn}) -- (\ref{outhellast}) with $r$ replaced by
$\chi$ and $\partial_\varphi$ replaced by the expressions in
Eq.~(\ref{FDMoutbc}).  For the fields $\widetilde{\Psi}^{(nn)}$,
$\widetilde{\Psi}^{(00)}$, $\widetilde{\Psi}^{(n0)}$ and
$\widetilde{\Psi}^{(20)}$, the conditions are just those used for the
scalar field in Ref.~\cite{eigenspec}.  For the complex fields, the
conditions are a modification of Eqs.~(\ref{outbcU}) and
(\ref{outbcV}), and are the tensorial equivalent of the adapted-coordinate
outer boundary condition (\ref{FDMoutbcapprox}) used for scalar
fields; aside from corrections of order $(\chi/a)^2$ these results
are:
\begin{eqnarray}
\frac{1}{\chi}\,\frac{\partial}{\partial\chi}\left(\chi U^{(ab)}
\right)=\pm\Omega\left(
-kV^{(ab)}+
\Gamma^\Theta\frac{\partial{U}^{(ab)} }{\partial\Theta}
+\Gamma^\Phi\frac{\partial{U}^{(ab)} }{\partial\Phi}
+\Gamma^\chi\frac{\partial{U}^{(ab)} }{\partial\chi}
\right)\label{outbcUadap}  \\
\frac{1}{\chi}\,\frac{\partial}{\partial\chi}\left(\chi V^{(ab)}
\right)=\pm\Omega\left(
kU^{(ab)}+
\Gamma^\Theta\frac{\partial{V}^{(ab)} }{\partial\Theta}
+\Gamma^\Phi\frac{\partial{V}^{(ab)} }{\partial\Phi}
+\Gamma^\chi\frac{\partial{V}^{(ab)} }{\partial\chi}
\right)
\label{outbcVadap}\,,
\end{eqnarray}
where
\begin{equation}
k=1 \mbox{\ \ for $(ab)=(n1), (n0),  (21)$}
\quad\quad
k=2 \mbox{\ \ for $(ab)=(22)$}\ .
\end{equation}

\section{Numerical results}\label{sec:numresults}

The field equations in adapted coordinates for
$\widetilde{\Psi}^{(nn)}$, $\widetilde{\Psi}^{(00)}$,
$\widetilde{\Psi}^{(n0)}$ and $\widetilde{\Psi}^{(20)}$ are defined by
the homogeneous wave equation~(\ref{BoxWTPsi00}) and by the form of the
d'Alembertian in Eq.~(\ref{waveq}) with the coefficients in Appendix
\ref{app:3Dcoeffs}.  These equations are subject to the inner boundary
conditions in Eqs.~(\ref{innerfirstadap}) -- (\ref{inner20adap})
applied at some smallest value $\chi_{\rm min}$ of $\chi $. The
equations must satisfy the outgoing or ingoing outer boundary
conditions Eq.~(\ref{FDMoutbcapprox}) applied at a largest
computational value $\chi_{\rm max}$ of $\chi$.  In the linear
theory each of the scalar-like fields
$\widetilde{\Psi}^{(nn)}$, $\widetilde{\Psi}^{(00)}$,
$\widetilde{\Psi}^{(00)}$, $\widetilde{\Psi}^{(20)}$ is completely
decoupled from every other fields, both in the field equations and in
the boundary conditions. Furthermore, the field equations, and
boundary conditions have {\rm precisely} the same forms as those for
the scalar field problem.  The
computational problem, therefore, is precisely that of
Ref.~\cite{eigenspec} where it was shown that the computed solution
agrees accurately with the numerically evaluated series solution in
Eq.~(\ref{allPsi}).

In the computation of the complex fields $\widetilde{\Psi}^{(n1)}$ and
$\widetilde{\Psi}^{(22)}$, features arise that are different from those
in Ref.~\cite{eigenspec}. Here the field equations (\ref{BoxU}) and (\ref{BoxV}),
and the outer boundary conditions (\ref{FDMoutbcapprox}),
couple the real and imaginary parts.
[This would apply also to the complex field $\widetilde{\Psi}^{(21)}$,
but due to the inner boundary condition $\widetilde{\Psi}^{(21)}=0$ in
Eq.~(\ref{innerlastadap}), the field $\widetilde{\Psi}^{(21)}$ must be 
be identically zero in linearized theory.]

\begin{figure}[ht] %%%%%  FIG 4
\includegraphics[width=.4\textwidth]{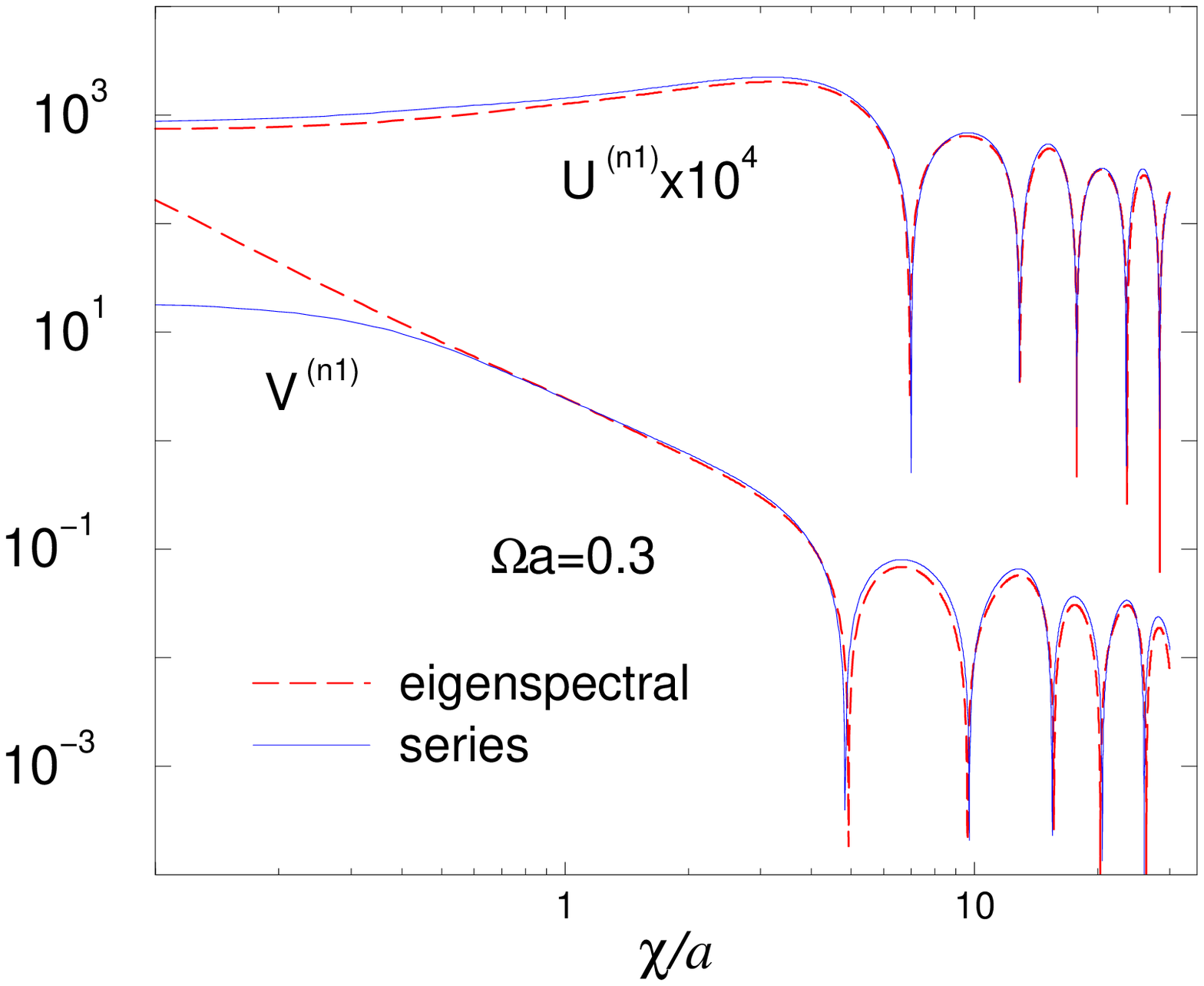}
\hspace{.3in}\includegraphics[width=.4\textwidth]{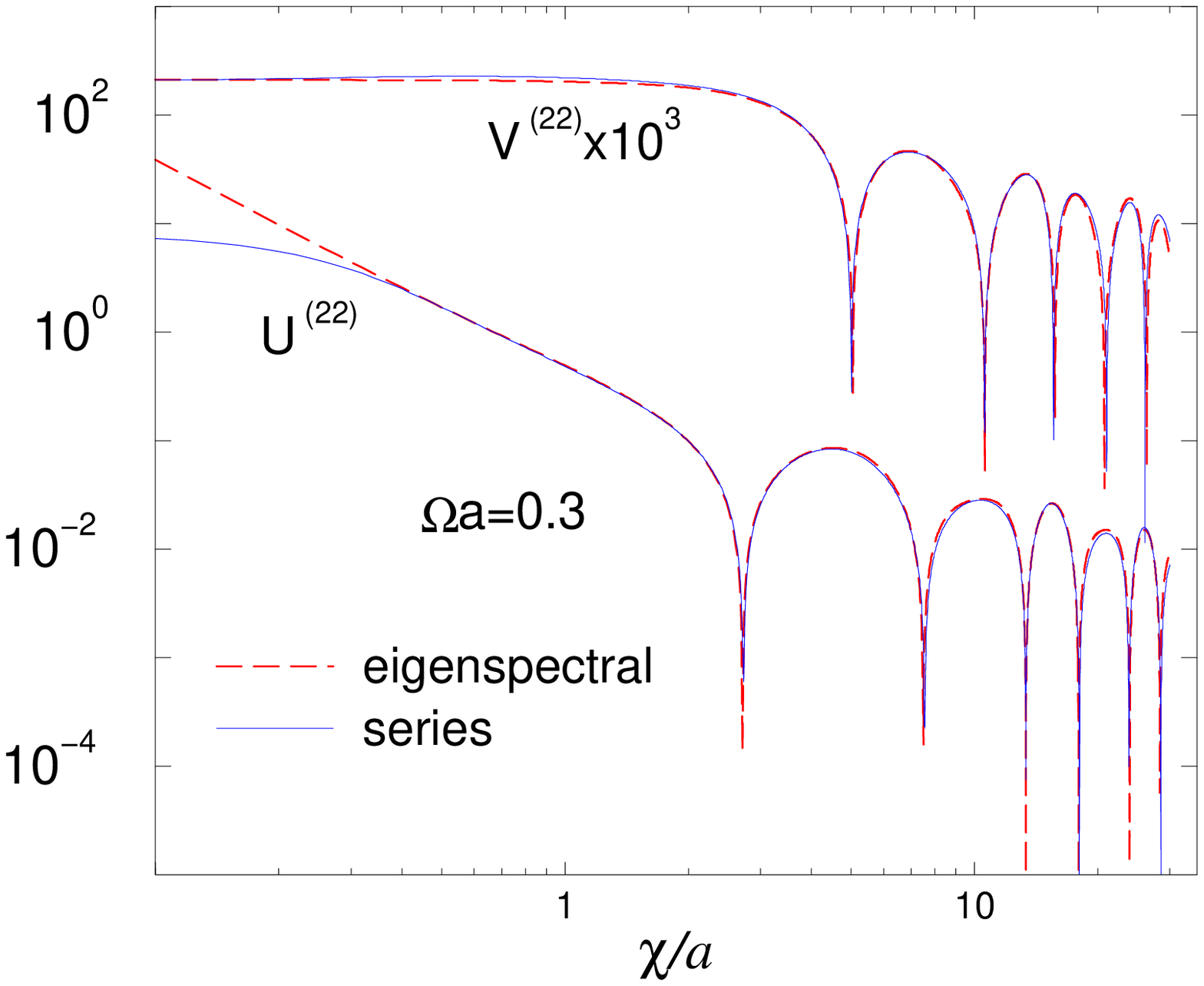}
\caption{ 
A comparison, for $a\Omega=0.3$, of outgoing linearized gravitational
fields computed by series summation (solid curve), and by the solution
of the coupled partial differential equations of the ``eigenspectral''
method of the PSW approach (dashed curve).  The fields are shown along
a line (the $\Theta=0$ line) outward through the source.
For these computations,
the grid in $\chi$, $\Theta $, $\Phi $ was 1500, 16, 32,
respectively. The entire angular space, $0\leq\Theta\leq\pi$,
$0\leq\Phi\leq2\pi $, was used. The inner boundary was at $\chi_{\rm
min}=0.1a$ and the outgoing boundary condition was imposed at
$\chi_{\rm max}=30a$.  The 
multipole filtering 
included all modes up through the octupole, approximately $\ell=3
$. (The actual discrete eigenvalues differ slightly
from integer values.) 
See the text for a discussion of the numerical limitations of 
the series summation.
\label{fig:Omega0.3}}
\end{figure}

\begin{figure}[ht]   %%%%FIG 5
\includegraphics[width=.4\textwidth]{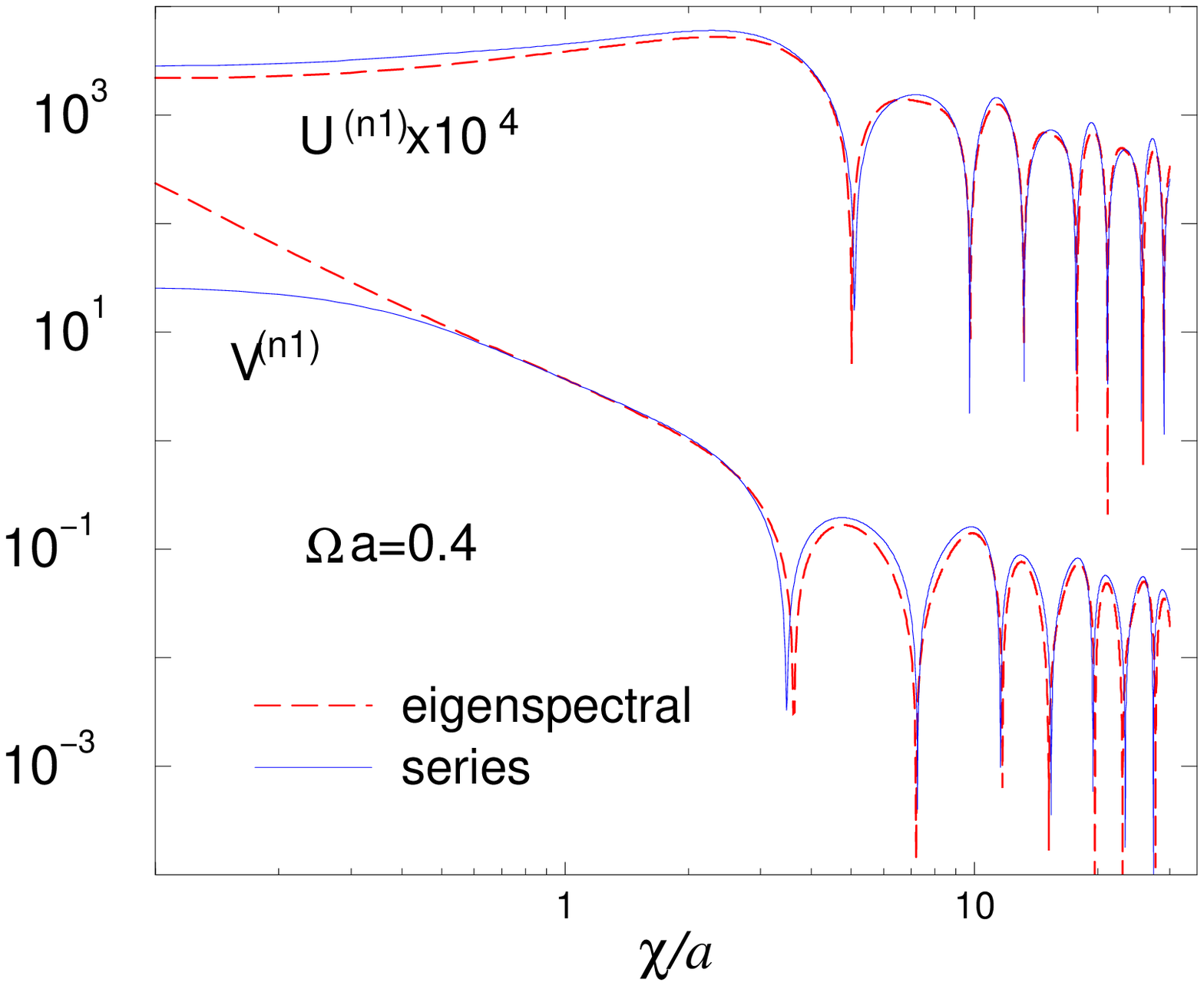}
\hspace{.3in}\includegraphics[width=.4\textwidth]{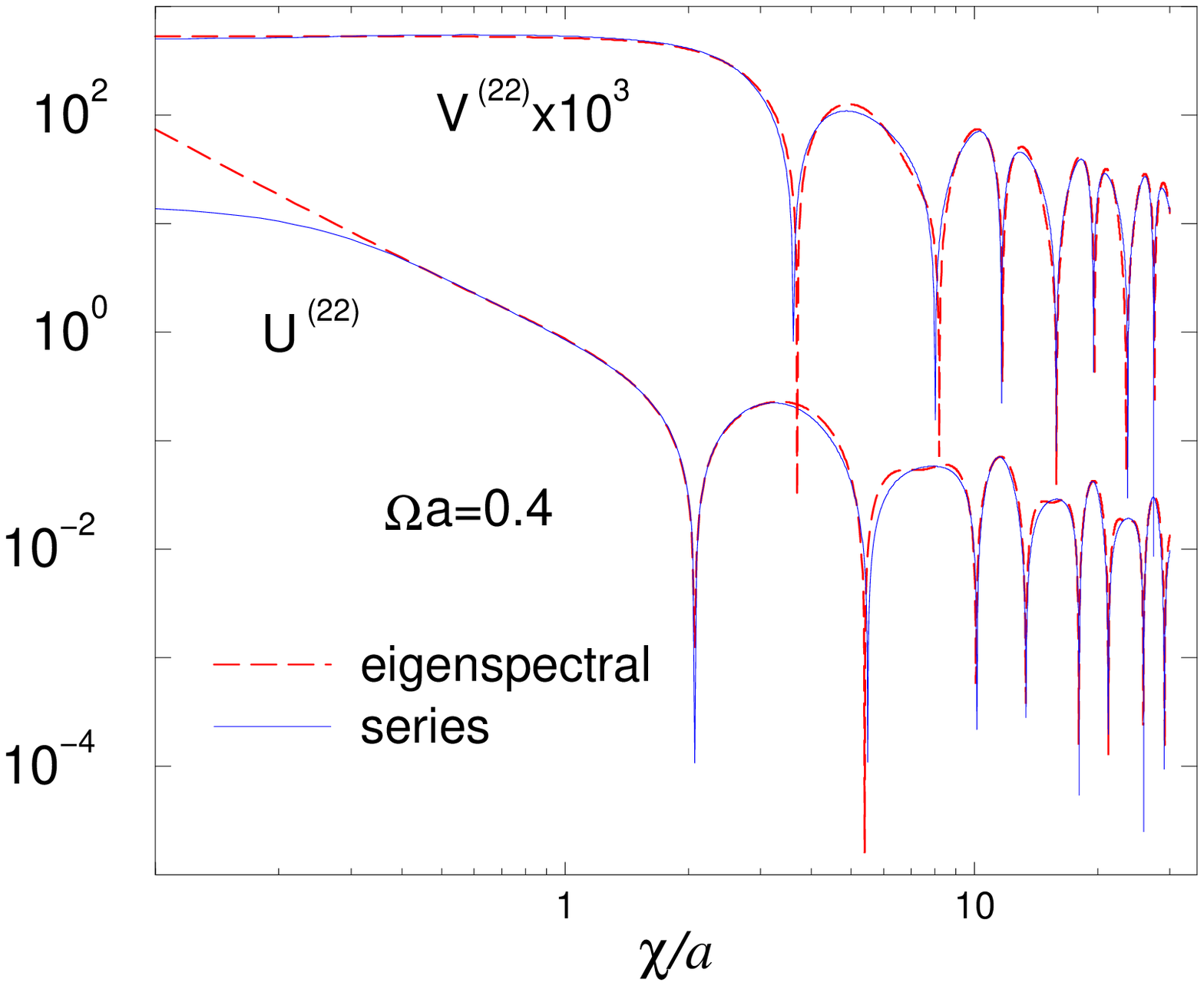}
\caption{
A comparison, for $a\Omega=0.4$, of outgoing linearized gravitational 
fields computed by  series summation (solid curve), and by
the solution of the coupled partial differential equations of 
the ``eigenspectral'' method of the PSW approach (dashed curve).
The details of the computations are the same as those for Fig.~\ref{fig:Omega0.3}.
\label{fig:Omega0.4}}
\end{figure}

The series solutions for ${U}^{(n1)}$ and ${V}^{(n1)}$ are given in
Eqs.~(\ref{Un1soln}) -- (\ref{Vn1soln}) and those for ${U}^{(22)}$ and
${V}^{(22)}$ in Eqs.~(\ref{U22soln}) -- (\ref{V22soln}).  Figures
\ref{fig:Omega0.3} and \ref{fig:Omega0.4} give a comparison, for two
different source velocities, between these series solutions and the
solutions of the eigenspectral method: adapted coordinates and
multipole filtering based on the modified multipoles appropriate to
the discrete angular operator.   

A feature that stands out in the figures is the disagreement at small
$\chi$ between the series and eigenspectral solutions for $V^{(n1)}$
and for $U^{(22)}$. The failing here is in the convergence of the
series solutions in Eqs.~(\ref{Vn1soln}) and (\ref{U22soln}).  The
$V^{(n1)}$ and $U^{(22)} $ fields diverge at $\chi\rightarrow0$, hence
these series converge very slowly at small $\chi$. Numerical
experiments summing very large numbers of terms and evalutating them
with arbitrary precision arithmetic confirm that the series solutions
for $V^{(n1)}$ and $U^{(22)}$ in Figs.~\ref{fig:Omega0.3} and
\ref{fig:Omega0.4} have large errors.  The series for $U^{(n1)}$ and
$V^{(22)}$, on the other hand, are convergent at $\chi=0$ and show
good agreement with the small-$\chi$ form of the eigenspectral
solutions.

The series solutions are highly accurate for $\chi>1$,
so the differences between the series solutions and the eigenspectral
solutions for large $\chi$ are an indication of the limitations of the
eigenspectral method. Those differences are larger for $\Omega>0.4$
than for $\Omega>0.3 $. Computations (not presented here) for $\Omega>0.5$ show
significantly larger error. The origin of these errors is the
relatively coarse computational grid used, and the limited number of
multipoles used in the multipole filtering. (This is equivalent in the
computation to the coarseness of the angular grid.) The error is simply
that due to truncation error, and is expected.
As $\Omega
$ is increased the gradients of the fields increase and truncation error
becomes more important.
These limitations
are imposed by the fact that the computation was done on a 2GB RAM
workstation. Greater accuracy, and hence higher velocity, 
would be possible on larger machines,

%%%%%%%%%%%%%%%%%%%%%%%%%%%%%%%%%%%%%%%%%%%%%%
%%%%%%%%%%%%%%%%%%%%%%%%%%%%%%%%%%%%%%%%%%%%%%
%%%%%%%%%%%%%%%%%%%%%%%%%%%%%%%%%%%%%%%%%%%%%%
%%%%%%%%%%%%%%%%%%%%%%%%%%%%%%%%%%%%%%%%%%%%%%

To test whether our numerical techniques might be sufficient for 
the next steps in our program, we introduced a simple nonlinearity
into the equations of Sec.~\ref{sec:adapcoord}

\begin{equation}\label{quadnonlin} 
\Box\bar{h}^{nn}
=\Box\widetilde{\Psi}^{(nn)}
=\kappa \frac{S}
{H^2
+a^2S}
\end{equation}
with $S
$  defined by
\begin{equation}
S=\eta^{\lambda\sigma}\bar{h}^{n\alpha}_{,\lambda}\bar{h}^{n\beta}_{,\sigma}\eta_{\alpha\beta}
\end{equation}
where $\bar{h}^{n\alpha}=({\bf n}\cdot\bar{\bf h})^\alpha$.

With the notation of Eqs.~(\ref{hbarinertial}) this can be written as
\begin{equation}
S=-\eta^{\lambda\sigma}\Psi^{(nn)}_{,\lambda}\Psi^{(nn)}_{,\sigma}
+\eta^{\lambda\sigma}\Psi^{(n0)}_{,\lambda}\Psi^{(n0)}_{,\sigma}
+\eta^{\lambda\sigma}\Psi^{(n1)}_{,\lambda}\left(\Psi^{(n1)}\right)^*_{,\sigma}\,.
\end{equation}
Since 
$\Psi^{(nn)}$ and $\Psi^{(n0)}$ are ``helical scalars,''
i.e.\,, functions only of corotating coordinates, for $(ab)=(nn)
$ or $(n0)$,
we  have 
\begin{displaymath}
\eta^{\lambda\sigma}\Psi^{(ab)}_{,\lambda}\Psi^{(ab)}_{,\sigma}
=\eta^{\lambda\sigma}\widetilde{\Psi}^{(ab)}_{,\lambda}
\widetilde{\Psi}^{(ab)}_{,\sigma}=\hspace{4in}
\end{displaymath}
\begin{equation}
-\Omega^2\widetilde{\Psi}^{(ab)}_{,\varphi}\widetilde{\Psi}^{(ab)}_{,\varphi}
+\left({\bf\nabla}\chi\cdot\nabla\chi\right)
\,\widetilde{\Psi}^{(ab)}_{,\chi}\widetilde{\Psi}^{(ab)}_{,\chi}
+\left({\bf\nabla}\Theta\cdot\nabla\Theta\right)
\,\widetilde{\Psi}^{(ab)}_{,\Theta}\widetilde{\Psi}^{(ab)}_{,\Theta}
+\left({\bf\nabla}\Phi\cdot\nabla\Phi\right)
\,\widetilde{\Psi}^{(ab)}_{,\Phi}\widetilde{\Psi}^{(ab)}_{,\Phi}\,,
\end{equation}
in which 
\begin{eqnarray}
\vec\nabla\chi\cdot\vec\nabla\chi&=&\frac{Q}{\chi^2}\\
\vec\nabla\Theta\cdot\vec\nabla\Theta&=&\frac{Q}{\chi^4}\\
\vec\nabla\Phi\cdot\vec\nabla\Phi&=&
2\;{\frac {Q+{a}^{2}+{\chi}^{2}\cos(2\,\Theta)}{{\chi}^{4}
 \sin^2(2\,\Theta)}}\label{delPhisq}\\
Q&\equiv&\sqrt{{a}^{4}+2\,{a}^{2}{\chi}^{2}\cos(2\Theta)+{\chi}^{4}}\label{Qdef} \ .
\end{eqnarray}

For $\Psi^{(n1)}$ we have 
$(\Psi^{(n1)}=e^{-i\Omega t}\widetilde{\Psi}^{(n1)}
$ 
which results in 
\begin{displaymath}
\eta^{\lambda\sigma}\Psi^{(n1)}_{,\lambda}\left(\Psi^{(n1)}_{,\sigma}\right)^*
=-\Omega^2
\widetilde{\Psi}^{(n1)}\left(\widetilde{\Psi}^{(n1)}\right)^*
-i\Omega^2\left[
\widetilde{\Psi}^{(n1)}\left(\widetilde{\Psi}^{(n1)}_{,\varphi}\right)^*
-\widetilde{\Psi}^{(n1)}_{,\varphi}
\left(\widetilde{\Psi}^{(n1)}\right)^*
\right]
-\Omega^2
\widetilde{\Psi}^{(n1)}_{,\varphi}\left(\widetilde{\Psi}^{(n1)}_{,\varphi}
\right)^*
%
%
%\eta^{\lambda\sigma}\widetilde{\Psi}^{(n1)}_{,\lambda}\widetilde{\Psi}^{(n1)}_{,\sigma}=\hspace{3in}
\end{displaymath}
\begin{displaymath}
+\left({\bf\nabla}\chi\cdot\nabla\chi\right)
\,\widetilde{\Psi}^{(n1)}_{,\chi} \left(\widetilde{\Psi}^{(n1)}_{,\chi}\right)^*
+\left({\bf\nabla}\Theta\cdot\nabla\Theta\right)
\,\widetilde{\Psi}^{(n1)}_{,\Theta}\left(\widetilde{\Psi}^{(n1)}_{,\Theta}\right)^*
+\left({\bf\nabla}\Phi\cdot\nabla\Phi\right)
\,\widetilde{\Psi}^{(n1)}_{,\Phi}\left(\widetilde{\Psi}^{(n1)}_{,\Phi}\right)^*
\end{displaymath}
\begin{equation}
=-\Omega^2
\left[\left({U}^{(n1)}\right)^2+\left({V}^{(n1)}\right)^2
\right]
+2\Omega^2
\left[
{V}^{(n1)}{U}^{(n1)}_{,\varphi}
-{U}^{(n1)}{V}^{(n1)}_{,\varphi}
\right]
-\Omega^2
\left[
\left({U}^{(n1)}_{,\varphi}\right)^2
+\left({V}^{(n1)}_{,\varphi}\right)^2
\right]
\end{equation}
\begin{displaymath}
+\left({\bf\nabla}\chi\cdot\nabla\chi\right)
\left[
\left({U}^{(n1)}_{,\chi}\right)^2
+\left({V}^{(n1)}_{,\chi}\right)^2
\right]
+\left({\bf\nabla}\Theta\cdot\nabla\Theta\right)
\left[
\left({U}^{(n1)}_{,\Theta}\right)^2
+\left({V}^{(n1)}_{,\Theta}\right)^2
\right]
+\left({\bf\nabla}\Phi\cdot\nabla\Phi\right)
\left[
\left({U}^{(n1)}_{,\Phi}\right)^2
+\left({V}^{(n1)}_{,\Phi}\right)^2
\right]\,.
\end{displaymath}
In the modified ``theory'' represented by Eq.~(\ref{quadnonlin}) the
equations that determine $\widetilde{\Psi}^{(n0)}$ and
$\widetilde{\Psi}^{(n1)}$ remain unchanged, so these fields are found
with linear equations.  The nonlinear occurrence of
$\widetilde{\Psi}^{(nn)} $ in $S$, however, means that
$\widetilde{\Psi}^{(nn)}$ solves a nonlinear equation.  

\begin{figure}[ht] %%%% FIG 6
\includegraphics[width=.4\textwidth]{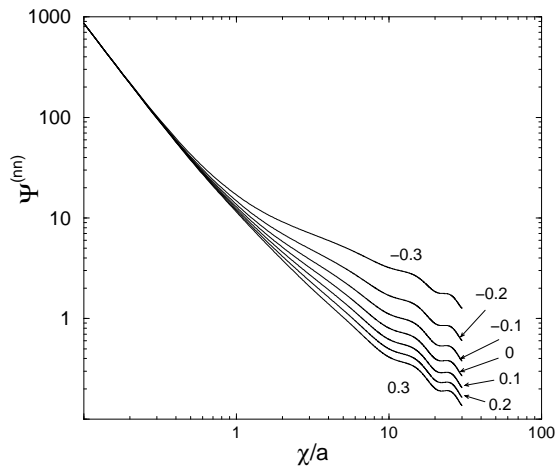}
\caption{The field $\widetilde{\Psi}^{(nn)}(\chi)
$ for the toy nonlinear model described in the text. 
Curves are marked with the value of the nonlinearity parameter $\kappa
$; the parameter $H
$ was set to unity and the models were run on a 
$1500\times16\times32
$ grid for the full $\chi,\Theta,\Phi
$ space (i.e.\,, no symmetries were used), with $\chi_{\rm min}/a=0.1$,
and $\chi_{\rm max}/a=30$.  In the multipole filtering, modes through 
octupole were included for all fields.
Comparison
with the $\kappa=0
$ linear solution shows the importance of nonlinear effects.
\label{fig:nonlin}}
\end{figure}

The solution
to this nonlinear problem requires the iterative techniques that have
previously been used for nonlinear scalar models in
Ref.~\cite{eigenspec}. Results from the applications of these
iterative methods to the modified theory are displayed in
Fig.~\ref{fig:nonlin} for various values of the nonlinearity parameter
$\kappa $. For all values of $\kappa $ the inner boundary conditions
are taken to be those of the linear problem and are imposed at
$\chi_{\rm min}=0.1$; for all models outgoing boundary conditions are
imposed at $\chi_{\rm max}=30$.
A comparison, in that figure, with the $\kappa=0$ linear solution
demonstrates that the cross-coupling of fields and nonlinearity has a
strong effect on the solution to the toy model, changing the amplitude
of the $\widetilde{\Psi}^{(nn)} $ waves by an order of magnitude.  The
purpose of this nonlinear toy computation is not to extract physics,
but simply to suggest that the iteration techniques previously
developed will be adequate at least for a range of nonlinear models.

%%%%%%%%%%%%%%%%%%%%%%%%%%%%%%%%%%%%%%%%%%%%%%
%%%%%%%%%%%%%%%%%%%%%%%%%%%%%%%%%%%%%%%%%%%%%%

%
\section{Summary and discussion}\label{sec:conc}

This paper has laid out the infrastructure for describing helically
symmetric linearized gravity, and more generally for describing
helically symmetric second-rank tensors in a flat background.  The
fields have been written in terms of ``helical scalars,'' i.e.\,,
functions only of corotating coordinates (equivalently, labels on the
curves of the Killing congruence).  The paper has, furthermore, shown
how to formulate computational problems in linearized general
relativity in terms of these helical scalars. The field equations,
sources, and inner and outer boundary conditions have been written in
this format in Sec.~\ref{sec:descrip}.

A welcome feature of the formulation in Sec.~\ref{sec:descrip} is the
extent of the separation of ``components'' of the field.  In the
starting point, the Lorentz-gauge field equations 
$\bar{h}_{\mu\nu,\alpha}^{\ \ \ \ ,\alpha}=-16\pi T_{\mu\nu}$, a
separate equation is satisfied by each component
$\bar{h}_{\mu\nu}$ with respect to an inertial Minkowski basis.
That attractive feature cannot be taken over directly to the helically
symmetric problem since the components with respect to the
inertial basis are not helical scalars. It turns out, however, that
the formulation of the linear problem in helical scalars leads to four
real fields $\tilde{\Psi}^{(nn)}$, $\tilde{\Psi}^{(n0)}$,
$\tilde{\Psi}^{(00)}$ and $\tilde{\Psi}^{(20)}$, and three complex
fields $\tilde{\Psi}^{(n1)}$, $\tilde{\Psi}^{(21)}$ and
$\tilde{\Psi}^{(22)}$, and that these four real and three complex
fields are not mixed by the field equations or the inner or outer
boundary conditions. The only mixing in the problem is between the
real and imaginary parts of the complex fields.

Since the three corotating
coordinates are general, this
paper has also presented the explicit formulation of the computational
problem in the adapted coordinates that were found in
Ref.~\cite{eigenspec} to be extremely useful. 
The numerical results in
Sec.~\ref{sec:numresults} of the present paper demonstrate that the
numerical challenges presented by linearized gravity are the same as
those for the linear scalar model in Ref.~\cite{eigenspec}. 
Indeed, the numerical problem for the four real fields is {\em exactly}
the same as that for the linear scalar field. For the complex 
fields, the new features are minor modifications of the boundary 
conditions and mixing of the real part and of the imaginary part
of each of the complex fields.  These new features do not appear
to present any new numerical difficulty, and in fact no difficulty
was found. Furthermore, a trial
with a toy nonlinearity suggests that there are also no new
problems in dealing with nonlinear terms, except those of complexity.

Although the present paper deals almost exclusively with linearized
gravity, the infrastructure developed here is more widely
applicable. Our next step in the PSW program is to solve the
post-Minkowskian problem for the orbiting point masses.  The equations
to be solved in this method have the same operator
$\bar{h}_{\mu\nu,\alpha}^{\ \ \ \ ,\alpha}$ on the left hand side, but
have ``sources'' quadratic in $\bar{h}_{\mu\nu,\alpha}$ on the
right. (There is also a second derivative on the right multiplied by
an undifferentiated $\bar{h}_{\mu\nu}$; this term can be treated, like
the others, as a perturbation, or it can be moved to the left to
modify the principal part.)  In the usual spirit of a post-Minkowski
approximation, we could solve first for the first-order fields and
treat them as known sources.  Alternatively, in a numerical approach,
we could treat the equations as a given nonlinear problem. Either way,
the formalism developed in the present paper goes over directly to the
post-Minkowskian problem. Again, the fields can be described with four
real and three complex helical scalars, and the equations and boundary
conditions follow from the simple relations between the helical
scalars and their inertial equivalents, i.e.\,, relations like those
in Eqs.~(\ref{tildePsifirst}) -- (\ref{tildePsilast}).

The full Einstein equations can also be viewed as a higher order
extension of the post-Minkowski equations. In principle, the only
change from the post-Minkowski problem is the inclusion of terms of
all orders on the nonlinear right-hand side of the field equations.
Again the formalism developed in the present paper should suffice for
the description of the problem, and should be convenient. In
particular, helical scalars of the background Minkowski space will be
helical scalars of the full metric.  In practice, new problems will
arise. One is the question of the relativistic Kepler's law: what is
the appropriate relationship relating the source strength (encoded in
inner boundary conditions), the coordinate separation of the sources,
and the parameter $\Omega $?  The post-Minkowskian approximation, for
which the answer is known, will help clarify how this is to be handled
in full general relativity. A more subtle question is whether it is
justified to use a formalism based on weak-field structures to
describe strong gravitational fields. For highly curved spacetimes do
Minkowksi-like coordinates exist with which we can use the formalism
of the present paper?  Possibly relevant to this is the fact that in
our computations we can impose inner boundary condistions at some
distance from the sources, so that the effects of strong fields can be
somewhat controlled.

In any case, the relative simplicity of the description presented
here, along with the absence of any new computational difficulties (that
is, difficulties not present in the scalar problem) is a reason for
optimism that the next steps can be taken reasonably quickly.

\section{Acknowledgment} 
We gratefully acknowledge the support of NSF grant PHY0244605 and NASA
grant ATP03-0001-0027, to UTB\@.
We thank Kip Thorne, Lee Lindblom, Mark Scheel and the Caltech numerical 
relativity group, and John Friedman.

%
%

%%%%%%%%%%%%%%%%%%%%%%%appendices %%%%%%%%%%%%%%%%%%%%%%%%%%%%%%%%%%%%%

\appendix

\section{Coefficients for adapted
coordinates}\label{app:3Dcoeffs}

The adapted-coordinate  coefficient $A$, $B$, and $\Gamma
$ are listed here. Derivations are provided in Ref.~\cite{eigenspec}.

As in Ref.~\cite{eigenspec} the coefficients needed for the wave operator
in Eq.~(\ref{waveq}) are written in the form:
\begin{eqnarray}
A_{\chi\chi}&=&\frac{Q}{\chi^2}
-{\Omega^2}\bar{A}_{\chi\chi} \\
A_{\Theta\Theta}&=&\frac{Q}{\chi^4}
-{\Omega^2}\bar{A}_{\Theta\Theta}\\
A_{\Phi\Phi}&=&2\;{\frac {Q+{a}^{2}+{\chi}^{2}\cos(2\,\Theta)}{{\chi}^{4}
 \sin^2(2\,\Theta)}}
-{\Omega^2}\bar{A}_{\Phi\Phi}\\
A_{\chi\Theta}&=&-{\Omega^2}\bar{A}_{\chi\Theta}\\
A_{\chi\Phi}&=&-{\Omega^2}
\bar{A}_{\chi\Phi}\\
A_{\Theta\Phi}&=&-{\Omega^2}
\bar{A}_{\Theta\Phi}\\
B_{\chi}&=&\frac{a^2+2Q}{\chi^3}
-{\Omega^2}\bar{B}_{\chi}\\
B_{\Theta}&=&{\frac {\sqrt {Q+{a}^{2}
+{\chi}^{2}\cos(2\,\Theta)}\;}{\sqrt {Q-{a}^{2}-{\chi}^{2}\cos(2\,\Theta)}}}
\;\frac{\left (Q-{a}^2\right )}{\chi^4}
-{\Omega^2}\bar{B}_{\Theta}\\
B_{\Phi}&=&
-{\Omega^2}\bar{B}_{\Phi}\,,\label{BPhi}
\end{eqnarray}
where $Q$ is given in Eq.~(\ref{Qdef}).

The expressions multiplied by $\Omega^2
$ are:
\begin{equation}\label{barAchichi} 
\bar{A}_{\chi\chi}=
\frac{a^4\sin^2(2\Theta)\;\cos^2\Phi}{\chi^2}
\end{equation}
\begin{equation}
\bar{A}_{\Theta\Theta}=
\frac{\cos^2\Phi\;\left[\chi^2+a^2\cos(2\Theta)\right]^2}{\chi^4}
\end{equation}
\begin{equation}
\bar{A}_{\Phi\Phi}=
\sin^2\Phi\;\frac{
Q+a^2+\chi^2\cos(2\Theta)
}{
Q-a^2-\chi^2\cos(2\Theta)
}
\end{equation}
\begin{equation}
\bar{A}_{\chi\Theta}=
\frac{a^2\;\left[
\chi^2+a^2\cos(2\Theta)
\right]\;\sin(2\Theta)\cos^2\Phi}{\chi^3}
\end{equation}
\begin{equation}
\bar{A}_{\chi\Phi}=
-\frac{a^2\;\left[Q+a^2+
\chi^2\cos(2\Theta)
\right]\;\sin\Phi\cos\Phi}{\chi^3}
\end{equation}
\begin{equation}
\bar{A}_{\Theta\Phi}=
-{\frac {\sin(\Phi)\cos(\Phi)\left[ {a}^{2}+{\chi}^{2}\cos(2\,\Theta)+
Q\right]\left[{\chi}^{2}+{a}^{2}\cos(2\,\Theta)\right]}{{\chi}^{4}
\sin(2\,\Theta)}}
\end{equation}
\begin{equation}
\bar{B}_\chi=\frac{a^2\left[
\cos^2(\Phi)\left\{
3a^2\cos^2(2\Theta)-Q-2a^2+\chi^2\cos(2\Theta)
\right\}+Q+a^2+\chi^2\cos(2\Theta)
\right]}{\chi^3}
\end{equation}
\begin{equation}\label{barBPhi} 
\bar{B}_\Phi=\frac{\left(3Q+a^2+\chi^{2}\cos{2\Theta}\right)
\sin(\Phi)\cos(\Phi)}
{Q-a^2-\chi^{2}\cos{2\Theta}}
\end{equation}
\begin{equation}
\bar{B}_\Theta=\frac{
\sqrt{Q+a^2+\chi^2\cos(2\Theta)}
}{\chi^6\sqrt{Q-a^2-\chi^2\cos(2\Theta)}
}
\; (c\cos^2\Phi+d)
\end{equation}
where
\begin{equation}\label{ceq} 
c\equiv {a}^{2}{\chi}^{4}\cos(2\,\Theta)+2\,{a}^{4}{\chi}^{2}+4\,{a}^{6}\cos(2
\,\Theta)+4\,{a}^{4}{\chi}^{2}\left (\cos(2\,\Theta)\right )^{2}-4\,{a
}^{4}Q\cos(2\,\Theta)-2\,{a}^{2}Q{\chi}^{2}-{\chi}^{6}
\end{equation}
\begin{equation}\label{deq} 
d\equiv{\chi}^{4}\left ({a}^{2}\cos(2\,\Theta)+{\chi}^{2}\right )\,.
\end{equation}

The coefficients needed in Eq.~(\ref{FDMoutbc}) a to express
the Sommerfeld boundary condition in adapted coordinates are:
\begin{equation}
\Gamma^\chi=
\left(\widetilde{Z}\frac{\partial\chi}{\partial\widetilde{X}}
-\widetilde{X}\frac{\partial\chi}{\partial\widetilde{Z}}\right)
=\frac{a^2\cos\Phi\,\sin(2\Theta)
}{\chi}
\end{equation}
\begin{equation}
\Gamma^\Theta=\left(\widetilde{Z}\frac{\partial\Theta}{\partial\widetilde{X}}
-\widetilde{X}\frac{\partial\Theta}{\partial\widetilde{Z}}\right)
=\frac{\cos\Phi\left(a^2\cos(2\Theta)+\chi^2\right)
}{\chi^2}
\end{equation}
\begin{equation}
\Gamma^\Phi=\left(\widetilde{Z}\frac{\partial\Phi}{\partial\widetilde{X}}
-\widetilde{X}\frac{\partial\Phi}{\partial\widetilde{Z}}\right)
=-\,\frac{\chi^2\sin\Phi\sin(2\Theta)}{-a^2-\chi^2\cos{2\Theta}+Q}
\end{equation}
Note that there were errors in the expressions given for 
the $\Gamma$s in Ref.~\cite{eigenspec}.


\begin{thebibliography}{10}

\bibitem{utbpuncture}
M.~Campanelli, C.~O.~Lousto, P.~Marronetti, and Y.~Zlochower,
preprint 
gr-qc/0511048; M.~Campanelli, C.~O.~Lousto, and Y.~Zlochower,
preprint 
gr-qc/0601091.





\bibitem{goddardpuncture} 
J.~G.~Baker, J.~Centrella, D.-I.~Choi, M.~Koppitz, and J.~van Meter,
preprint gr-qc/0505100.

\bibitem{laz1} 
J.~Baker, M.~Campanelli, C.~O.~Lousto, and
   R. Takahashi, 
\prd {\bf 65},  124012  (2002).

\bibitem{PN} L.~Blanchet, 
Living Rev. Relativity 5, (2002), 3. [Online article]: cited Dec.~25,
2000 http://www.livingreviews.org/lrr-2002-3.

\bibitem{det}
J.~K. Blackburn and S. Detweiler, Phys. Rev. D {\bf 46},  2318  (1992).

\bibitem{det2}
S. Detweiler, Phys. Rev. D {\bf 50},  4929  (1994).

\bibitem{WKP}
J.~T. Whelan, W. Krivan, and R.~H. Price, Class. Quant. Grav. {\bf 17},  4895
  (2000).

\bibitem{WBLandP}
J.~T. Whelan, C. Beetle, W. Landry, and R.~H. Price, Class. Quant. Grav. {\bf
  19},  1285  (2002).

\bibitem{rightapprox}
R.~H. Price, Class. Quant. Grav. {\bf 21},  S281  (2004).

\bibitem{paperI}
Z. Andrade {\it et~al.}, Phys. Rev. D {\bf 70},  064001  (2004).

\bibitem{helicaldef}
M. Shibata, J. L. Friedman, K.~Uryu, and M. Shibata
\prd {\bf 65},  064035 (2002).
Erratum-ibid. D {\bf70} 129904 (2004).


\bibitem{eigenspec} 
B. Bromley, R. Owen and R.~H.~Price
\prd
{\bf71}, 104017 (2005).

\bibitem{MTW} C.~W.~Misner, K.~S.~Thorne and J.~A.~Wheeler,
{\em Gravitation} (W.~H.~Freeman, San Francisco, 1973).

\bibitem{torre} 
C.~G.~Torre, J.~Math.~Phys., {\bf44} 
6223-6232 (2003).

\bibitem{zerilli70} F.~J.~Zerilli, \prd {\bf2}, 2141 (1970), Eq.~(1).

\bibitem{kiprevmodphys}
K.~S.~Thorne \rmp {\bf 52}, 299 (1980), Sec.~IIE.


\bibitem{mathews} 
J.~Mathews, J.~Soc.~Ind.~Appl.~Math. {\bf10}, 768 (1962).



\bibitem{reggewheeler}T.~Regge and J.~A.~Wheeler, Phys.\
Rev. {\bf108}, 1063 (1957).


\bibitem{sandberg}  V.~D.~Sandberg, J.~Math.~Phys  {\bf19}, 2441 (1978).

\bibitem{goldbergetal} 
J. N. Goldberg, A. J. MacFarlane, E. T
Newman, F. Rohrlich and E. C. G. Sudarshan, 
J.~Math.~Phys  {\bf8}, 2155 (1967).

\end{thebibliography}
\end{document}